\documentclass[12pt]{article}

\usepackage{amsmath}
\usepackage{amsfonts}
\usepackage{amssymb}
\usepackage{amsthm}
\usepackage{bm}
\usepackage{caption}
\usepackage{cite}
\usepackage{color}
\usepackage{graphicx}
\usepackage{slashed}            
\usepackage{xspace}				
\usepackage{fancyhdr}
\usepackage{psfrag}
\usepackage{lmodern}
\usepackage{physics}
\usepackage{setspace}
\usepackage{bm}
\usepackage{booktabs}
\usepackage[affil-it]{authblk}
\usepackage[separate-uncertainty=true]{siunitx}
\usepackage{multirow}
\usepackage[dvipsnames,x11names]{xcolor} 
\usepackage{subfig}
\usepackage{float}
\usepackage{bmpsize}
\usepackage[shortlabels]{enumitem}
\usepackage{listings}
\usepackage{dcolumn}
\usepackage[version=4]{mhchem}
\usepackage{mathtools}
\usepackage{xpatch}
\usepackage[hypertexnames=false]{hyperref}		
\definecolor{lightlightgray}{rgb}{.85,.85,.85}
\numberwithin{equation}{section}
\theoremstyle{definition}

\theoremstyle{plain}

\let\originalleft\left
\let\originalright\right
\renewcommand{\left}{\mathopen{}\mathclose\bgroup\originalleft}
\renewcommand{\right}{\aftergroup\egroup\originalright}

\newcommand{\arXiv}[2]{\href{http://arxiv.org/pdf/#1}{{\tt #2/#1}}}
\newcommand{\arXivold}[1]{\href{http://arxiv.org/pdf/#1}{{\tt #1}}}

\renewcommand{\tilde}{\widetilde} 

\newcommand{\beq}{\begin{eqnarray}}
\newcommand{\eeq}{\end{eqnarray}}

\newcommand{\bag}{\begin{align}}
\newcommand{\eag}{\end{align}}

\begin{document}

\begin{titlepage}
	
\hfill DESY 19-236	
\vspace{10pt}
\begin{center} 
{\huge \bf Radion-Activated Higgs Mechanism} 
\end{center}
\vspace*{0.4cm} 

\begin{center} 
{\bf \  Cem Er\"{o}ncel$^{a,b}$, Jay Hubisz$^b$, Gabriele Rigo$^b$} 

 $^{a}$ {\it DESY, Notkestrasse 85, D-22607 Hamburg, Germany} \\
 
 \vspace*{0.1cm}
 
 $^{b}$ {\it Department of Physics, Syracuse University, Syracuse, NY  13244, USA} \\
 
 \vspace*{0.1cm}

\vspace*{0.1cm}
{\tt  
 \href{mailto:cem.eroncel@desy.de}{cem.eroncel@desy.de},
 \href{mailto:jhubisz@syr.edu}{jhubisz@syr.edu},  
 \href{mailto:garigo@syr.edu}{garigo@syr.edu}}

\end{center}

\vglue 0.3truecm

\begin{abstract}
We study multi-scalar models of radius stabilization, with an eye towards application to novel extra-dimensional models of symmetry breaking.  With inspiration from holography, we construct a multi-scalar effective potential that is a function of UV-brane values of the scalar fields, and that takes into account bulk gravitational backreaction.  We study extrema of this potential, and additionally provide a ``superpotential" method for generating static solutions for the extra-dimensional geometry.  We apply these methods to some simple models of the Higgs mechanism where the Higgs itself plays a non-trivial role in radius stabilization.  We conclude that mass mixing of the Higgs and radion is generic unless additional symmetries are imposed.  We focus on models with moderate gap between the electroweak and Kaluza-Klein scale, as required by phenomenological constraints.  We note that tuning of the Higgs mass relative to the KK scale is related to various classes of tuning of 5D parameters, with different resulting spectra and phenomenologies.\end{abstract}
\end{titlepage}

\setcounter{equation}{0}
\setcounter{footnote}{0}
\setcounter{section}{0}

\section{Introduction}
The mass of the Higgs is far smaller than dimensional analysis predicts given that there are so far no experimental signatures of new physics indicating new symmetries near the electroweak scale.  This suggests that there may be some degree of fine tuning of parameters to achieve consistency with observations.  Of course, the Higgs mass problem is not the only one of the Standard Model.  The SM offers no explanation of large fermion mass hierarchies, of dark matter, of the matter/anti-matter asymmetry, etc.   Strong dynamics, and/or field theories in higher-dimensional spacetimes offer resolutions or at least interesting reformulations of many of these problems, along with partially ameliorating the Higgs hierarchy problem.

The Higgs mass couples to singlet sectors in many extensions of the electroweak sector, which can put the the hierarchy problem into a new light~\cite{Graham:2015cka}.  The question is no longer ``Why is the Higgs light?"  Rather, the correct question is ``Why does a small Higgs mass coincide with the global (or a cosmologically metastable) minimum of the scalar potential?"  

Such scalars are generic features in models with additional compact spatial dimensions~\cite{Randall:1999ee}.  In these models, there are typically moduli that are gauge singlet states with non-trivial couplings to matter required by higher dimensional general covariance~\cite{Goldberger:1999uk,Goldberger:1999un}.  Stabilization of these moduli is typically non-trivial as the moduli can be interpreted as Goldstone bosons of spontaneously broken spacetime symmetries.  Providing a potential for them often requires new scalar degrees of freedom whose dynamics serve to explicitly break the spacetime symmetries, demoting the moduli to pseudo-Goldstone bosons.

In this work, we explore the relationship between extra-dimensional radius stabilization and the Higgs mass.  In particular, the modulus of Randall-Sundrum models (the ``radion") generally plays a crucial role in determining the value of the Higgs vacuum expectation value.  The same goes for the reverse: the Higgs VEV itself will backreact on the geometry, and feed into the total effective potential for the radion.  

This first motivates the development of a formalism for dealing with multi-scalar models of 5D RS model radius stabilization.  This is the focus of the first part of this work.  We show that even in cases of complicated bulk dynamics with many interacting scalar fields, the classical effective potential lives on the boundary.  Making the assumption that the 5D action is of Einstein-Hilbert form with minimal couplings to bulk scalar fields, we calculate the classical potential to all orders in the 5D gravitational coupling constant.  

The criteria for minimization of this effective potential determine aspects of the physics such as the Kaluza-Klein scale, and the masses of light resonances.  We study these criteria, and provide conditions for extremization of this potential that fully include geometric backreaction effects.

We then apply this formalism to a few illustrative examples of the Higgs mechanism in RS models.   Given the lack of signals for extra-dimensional resonances in collider experiments, we look for models which achieve moderate separation of scales between the electroweak scale and the scale associated with the IR brane $f/v \gtrsim {\mathcal O}(10)$.   The Higgs mass term is generally a function of the brane separation.  Thus, in order to achieve a light Higgs, the minimum of the modulus potential must coincide with the region where the Higgs mass is small.  We expect (and find) that tuning is involved in these cases.  For example, such a coincidence problem may require that the IR brane mistune (between the brane tension and the bulk cosmological constant) be adjusted to high degree in order to keep the Higgs light.  We explore the type and degree of tuning required in a few examples in which there is non-trivial interplay between electroweak symmetry breaking and radius stabilization.

As noted in~\cite{Amin:2019qrx}, fine tuning can have dramatic phenomenological consequences when parameters of the Standard Model are set by the dynamics of a modulus field.  These consequences influence both early universe cosmology and collider physics.  Cosmology can be dramatically changed by interplay of Higgs and modulus field, with oscillations of the coupled scalar system leading to repercussions for gravity waves, and constraints on inflationary models.  It is with this in mind that we choose models whose parameter space explore the full range of classes of mixed modulus-Higgs potential.  The specifics of these classes are expanded on in the introduction to Section~\ref{sec:radionexamples}.

The example models we study are as follows:
\begin{itemize}
\item Higgs on the IR brane, Goldberger-Wise scalar stabilizing field in the bulk;
\item Higgs in the bulk, with mass near the Breitenlohner-Freedman bound~\cite{Breitenlohner:1982jf}, stabilizing the geometry (studied previously in~\cite{Geller:2013cfa,Vecchi:2010em});
\item Both Higgs and Goldberger-Wise stabilizing field in the bulk.
\end{itemize}
For simplicity of presentation, these models are all studied to lowest non-trivial order in the gravitational backreaction, where neither the Higgs or Goldberger-Wise scalar develop VEVs that are comparable to the AdS curvature.  

An important result from these studies is that mass mixing between the Higgs and radion is completely generic.  Mass mixing occurs in different ways: through symmetry-allowed couplings between a Goldberger-Wise field and the Higgs, or from mixing of the Higgs with the 5D gravity sector through backreaction.\footnote{This is in addition to kinetic mixing, which can occur through couplings of the Higgs to either the 5D curvature, or to the extrinsic curvature of the IR brane~\cite{Giudice:2000av}.}  It occurs both when the Higgs is purely localized on the IR brane through couplings to the Goldberger-Wise field, or when the Higgs is in the bulk of the extra dimension.  This mixing is different in the various classes of Higgs-radion potential, and influences the spectrum of light scalar modes.

In Section~\ref{sec:genprop} we discuss general properties of multi-scalar stabilization, giving conditions under which the geometry is stabilized.  In Section~\ref{sec:radionexamples}, we explore the construction, phenomenology, and spectra of three different models where the Higgs is coupled to the extra-dimensional radion.  In Section~\ref{sec:cftinterpretation} we discuss a CFT interpretation of different multi-scalar stabilization models.  In Section~\ref{sec:conclusions} we conclude.  In Appendix~\ref{sec:derivativederivation}, we include derivations of some results in Section~\ref{sec:genprop}.  In Appendix~\ref{sec:superpotential}, we give a presentation of a ``superpotential" method to generate static geometries with multiple scalar fields.


\section{General Properties of the Radion Potential}
\label{sec:genprop}
Before analyzing particular models, we first study some general properties of radius stabilization when there is more than one 5D scalar field that has non-vanishing vacuum expectation value.  In such cases, the radion potential is affected by backreaction of all of these VEVs onto the geometry.  In what follows, we consider arbitrary backreaction onto the geometry, however we presume that the action is of Einstein-Hilbert form, with no higher curvature operators. 

The generic problem is one of $N$ real scalar fields minimally coupled to gravity in 5D space with negative cosmological constant $\Lambda_5 = -6k^2/\kappa^2$.  The bulk geometry can be described with metric
\beq
\dd{s^2} = e^{-2 A(y)} \dd{x_4^2} -\dd{y^2},
\eeq
where we have presumed flat 4D slices.  The space is cut off on both sides by branes at positions $y_0$ and $y_1$.

All dimensionful quantities are understood to be expressed in units of the AdS curvature:  $k =1$.  For perturbative control of the 5D gravity theory, the 5D Newton constant must be small:  $\kappa^2 =\frac{1}{2 M_5^3} \sim {\mathcal O}(1/10)$.  The action is:
\beq
\begin{aligned}
	S &= \int \dd[4]{x} \dd{y} \sqrt{g} \left[  \frac{1}{2} \sum_i (\partial_M \phi_i)^2 -V( \{ \phi_i \} ) -\frac{1}{2 \kappa^2}R \right] \\
	&\phantom{{}={}}- \int \dd[4]{x}  \sqrt{-g_0} V_0 (\{ \phi_i \})\eval_{y=y_0} -\int \dd[4]{x} \sqrt{-g_1} V_1 (\{ \phi_i \})\eval_{y=y_1},
\end{aligned}
\eeq
where the scalar potential includes the bulk cosmological constant term.

The bulk Einstein equations relate derivatives of $A(y)$ to the scalar fields in the following way:
\beq
\begin{gathered}
{A'}^2 = \frac{\kappa^2}{12} \sum_i \phi_i'^2 - \frac{\kappa^2}{6} V(\{ \phi_i \}), \\
A'' = \frac{\kappa^2}{3} \sum_i {\phi_i'}^2.
\label{ay_profile}
\end{gathered}
\eeq
The scalar curvature is given by
\beq
R = 20 A'^2 - 8 A''.
\eeq
In the absence of scalar field VEVs, or in the limit of small $\kappa^2$, the space is AdS.

Plugging the result for the scalar curvature, and for the scalar terms into the  action above, we find that the bulk portion is a derivative, and thus can be expressed as a boundary term:
\beq
S_\text{bulk} = \frac{2}{\kappa^2} \int \dd[4]{x} \dd{y}  \frac{\partial}{\partial y} \left[ e^{-4 A(y)} A'(y) \right] = \frac{2}{\kappa^2} \int \dd[4]{x} \left[  e^{-4 A(y_0)} A'(y_0) -  e^{- 4 A(y_1)} A'(y_1)\right].
\eeq
The factor of 2 is from integration over the circle.
We note that on the boundaries, there must be a jump in the derivative of the metric.  This can be seen most easily from the orbifold perspective, where on either side of an orbifold fixed point, you must have $A(y_{0,1})_+ = A(y_{0,1})_-$ and $A'(y_{0,1})_+ =  -A'(y_{0,1})_-$.  This leads to a term in the scalar curvature $R$ which is a delta function, as you must have $A''(y_{0,1}) \supseteq \pm 2 \delta(y_{0,1}-y) A'(y_{0,1})$.  Inclusion of the delta function term in the scalar curvature gives additional boundary contributions to the effective action.

Including all terms, the effective potential can now be expressed as a pure boundary term~\cite{Bellazzini:2013fga}:
\beq
\label{geneffpot}
V_\textnormal{eff} = e^{-4 A (y_0)} \left[ V_0 (\{ \phi_i (y_0) \})  -\frac{6}{\kappa^2} A' (y_0) \right] + e^{-4 A (y_1)} \left[ V_1 (\{ \phi_i (y_1) \}) + \frac{6}{\kappa^2} A' (y_1) \right].
\eeq
where, at the moment, none of the boundary conditions have been imposed, nor even the bulk scalar equations of motion, which are given by:
\beq
\label{genbulkeom}
\phi_i'' = 4 A' \phi_i' + \frac{\partial V}{\partial \phi_i}.
\eeq
The scalar boundary conditions are given (in these coordinates) as:
\beq
\label{genbcs}
\phi'_i \eval_{y_0,y_1} = \pm \frac{1}{2} \frac{ \partial V_{0,1} }{\partial \phi_i} \eval_{y_0,y_1}.
\eeq
The boundary Einstein equations (the metric junction conditions that match the geometry to the brane-localized stress-energy) are
\beq
A' \eval_{y_0,y_1} = \pm \frac{\kappa^2}{6} V_{0,1} \eval_{y_0,y_1},
\eeq
which correspond to the UV and IR terms in the effective potential vanishing individually.

One can associate the brane separation with the vacuum expectation value of the radion.  The metric boundary condition at $y_1$ can be thought of as setting the VEV of the radion.  That is, one varies $y_1$ until the IR brane contribution to the effective potential vanishes.  We show in the next Subsection that it is always the case that the minimum in the static effective potential corresponds to vanishing $V_\text{IR}$ after all scalar bulk and boundary equations of motion are imposed.\footnote{However, it curiously remains possible that extrema or saddles arise where neither the scalar or metric conditions are satisfied (and this occurs in some models~\cite{Eroncel:2018dkg}).  This may be due to nearby extrema which do not satisfy the ansatz of a Lorentz invariant background.}

At this point we can impose the scalar field equations of motion.  For a single scalar field, if we impose the bulk equations, the effective potential is then a function of the two remaining freedoms in the scalar field (its value and derivative, on, for example, the UV brane), and the position of the IR brane.  If one further imposes the scalar boundary conditions on both of the two branes, then only the value of $y_1$ remains, to be determined by minimizing the effective potential with respect to it.  However, we might be interested in a more intuitive measure of the effective potential, particularly when taking into account the multiple scalar degrees of freedom, where we expect to have a multidimensional scalar potential.

We advocate an approach inspired by holography, where the effective potential is measured in terms of field values on the UV brane only.  In this case, the procedure we should take to find the effective potential is as follows:
\begin{itemize}
\item pick a particular value of $y_1$;
\item impose the IR boundary conditions, eliminating $N$ of the $2N$ scalar boundary conditions.  $N$ degrees of freedom remain, e.g. the values of the fields on the UV brane;
\item we are left with $N$ scalar boundary values and the free value of $y_1$.  Ideally, one would then, for constant UV field values, minimize the effective action over $y_1$.
\end{itemize}
This last step is difficult to implement in practice.  However, if at least one of the scalar fields (say $\phi_N$) has ``stiff" boundary conditions in the UV, then one instead imposes that UV brane condition, leaving $N-1$ scalar degrees of freedom on the UV brane.  The value of $y_1$ corresponds to the last degree of freedom, so that one has $V_\text{eff} (\phi_1(y_0),\cdots,\phi_{N-1} (y_0); y_1)$.


\subsection{Derivative of the Radion Potential}
We are interested in finding stable configurations of the branes, where the effective potential is minimized.  We explore here the conditions for minimization, showing analytically that they correspond in most, but not all, cases to vanishing of the IR portion of the effective potential.

Employing the above procedure, imposing the IR brane boundary conditions for the scalar fields the derivative of the effective potential with respect to $y_1$ takes a particularly simple form:
\begin{equation}
\dv{V}{y_1} = e^{-4 A(y_0)} \sum_i \left( \frac{\partial V_0}{\partial \phi_i}(y_0)-2 \phi_i' (y_0) \right) \dv {\phi_i}{y_1}\left(y_0\right) - 4 e^{-4 A_1} \dv{A_1}{y_1} \left[ V_1 + \frac{6}{\kappa^2} A'_1 \right].
\label{eq:dVbydy1}
\end{equation}
We provide proof of this relation in Appendix~\ref{sec:derivativederivation}.
The first term is vanishing if the boundary conditions for the scalars on the UV brane are imposed.  The second term is vanishing if the metric junction condition in the IR is imposed.  We thus note that this formula implies that extremization of the action implies (as it must, of course) extremization of the effective potential.  However, the converse is not necessarily the case.  It may be that this expression vanishes where the two terms cancel against each other and neither metric nor scalar boundary conditions are met.    We hypothesize that this may occur when there are extrema of the action ``nearby" in configuration space that do not obey the static, homogeneous, and isotropic ansatz that was the starting point for this analysis of the effective action.

We note that there is a method that generalizes the well known superpotential method for stabilization~\cite{DeWolfe:1999cp} to include multiple scalar fields.  This method guarantees a solution in which the effective potential is extremized.  In Appendix~\ref{sec:superpotential}, we derive and present this method.


\section{Examples of Radion-Induced Symmetry Breaking}
\label{sec:radionexamples}
\begin{figure}
\centering
	\subfloat{\includegraphics[width=0.3\textwidth,keepaspectratio]{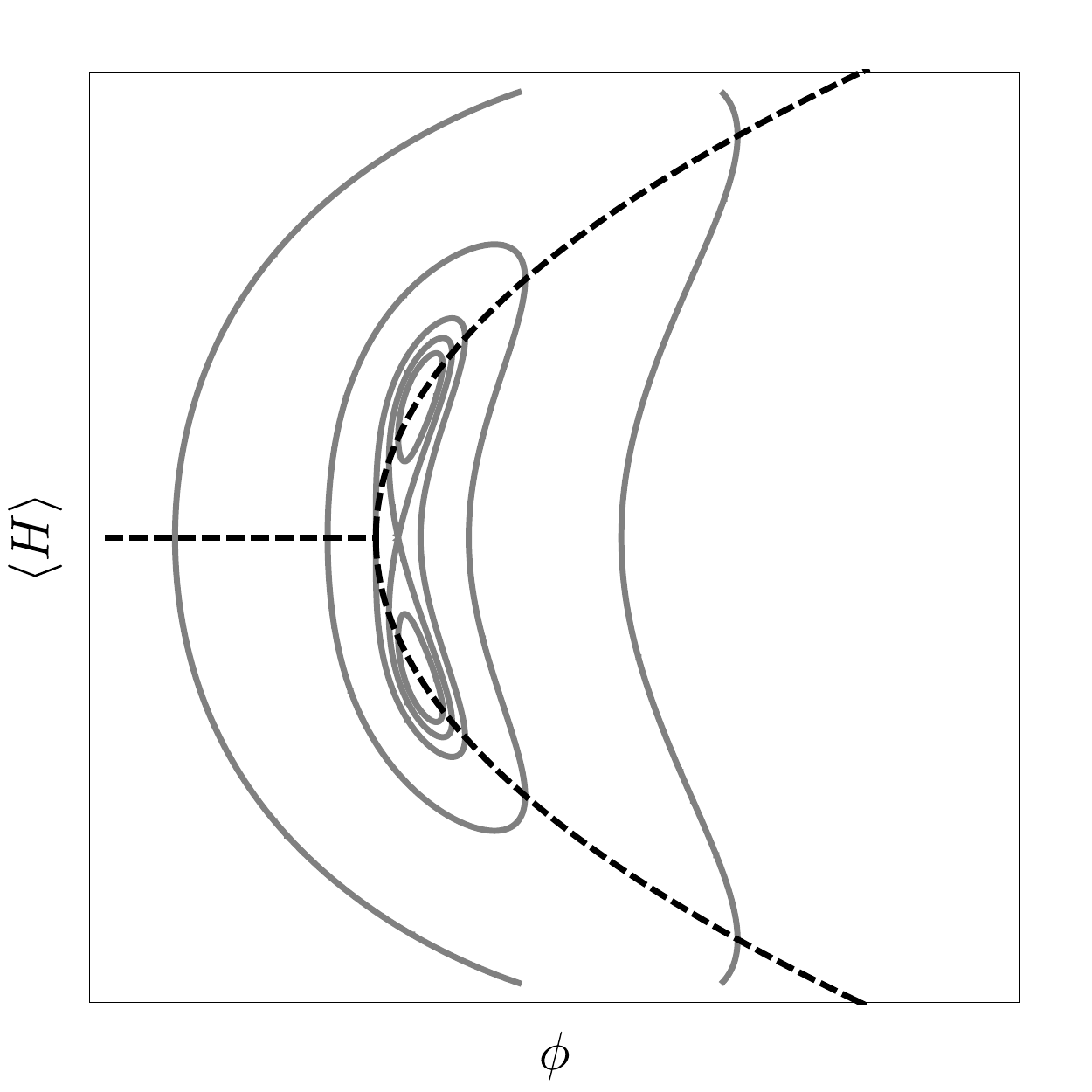}}\hfill%
	\subfloat{\includegraphics[width=0.3\textwidth,keepaspectratio]{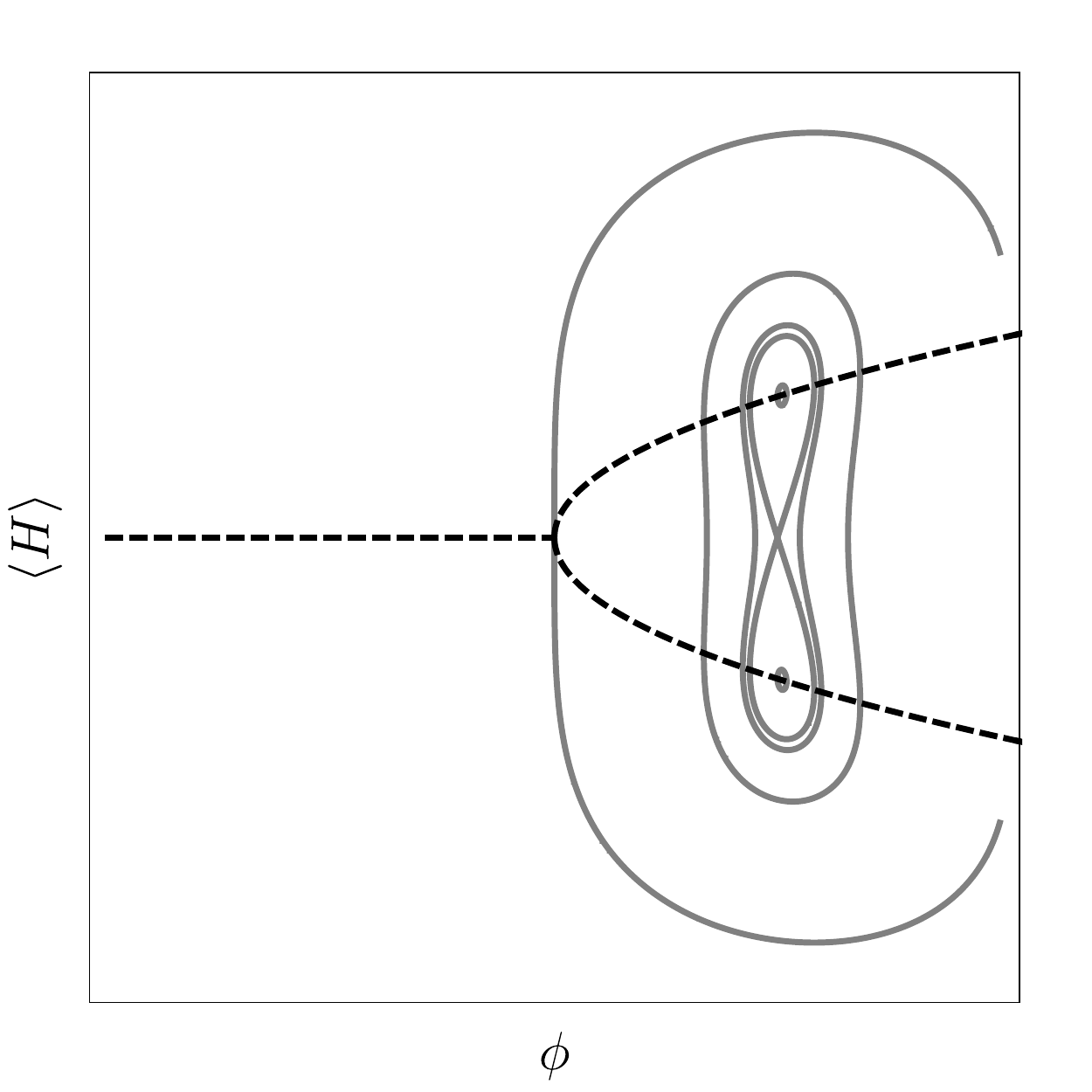}}\hfill%
	\subfloat{\includegraphics[width=0.3\textwidth,keepaspectratio]{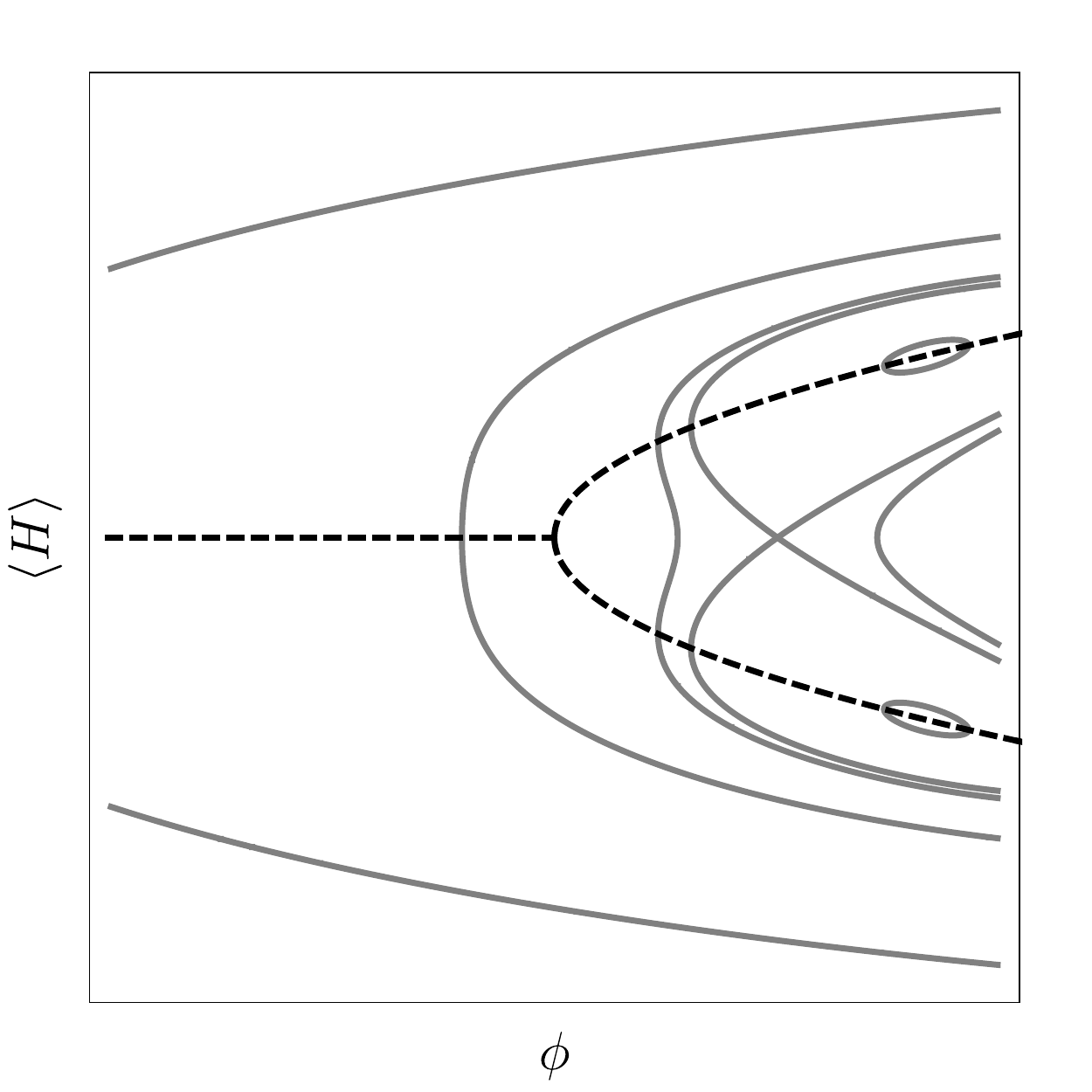}}
	\caption{In this figure, we roughly characterize the different types of potentials in the space of the radion and the Higgs VEVs.  Each plot shows a potential as a function of a modulus VEV, $\phi$, and a Higgs VEV.  Dashed lines indicate points at which the Higgs potential is minimized with $\phi$ held constant.  In the first image, all couplings are order 1, however, the minimum of the modulus potential is tuned to be close to the critical point for the Higgs.  In the second, the dependence of the Higgs potential on the modulus field is weak, and the bare Higgs mass is taken to be somewhat small.  In the final plot, the modulus potential without the Higgs is very flat, and thus the backreaction of the Higgs onto the modulus potential is the dominant feature in the potential.}
	\label{fig:cartoons}
\end{figure}

We now explore some specific simple models that illustrate different classes of multi-scalar potentials where a Higgs instability occurs over some region of  radius values.  We are particularly interested in models where some tuning has been performed to obtain mild or large hierarchies between the electroweak scale and the KK scale.  Broadly speaking, there are 3 classes of tunings distinguished by the shape of the 2-dimensional scalar potential.  These are as follows:
\begin{itemize}
\item All parameters of the 5D theory are of order 1 (in units of the AdS curvature for dimensionful couplings).  Tuning is achieved by making the symmetry breaking critical point extremely close to the minimum of the radion potential.
\item The Higgs potential has only mild dependence on the radion VEV.  Tuning is achieved by arranging the Higgs effective mass term to be small over a large range of radius, including at the minimum of the radion potential.
\item The radion potential itself is very shallow due to tuning.  The Higgs VEV then contributes sizably to the stabilization, and even small Higgs VEV successfully stabilizes the geometry for large KK scale.
\end{itemize}
In Figure~\ref{fig:cartoons}, we display the basic pictures of these three types of 2D scalar potentials.

These different types of tunings have implications for phenomenology.  For example, there may be consequences for cosmology, affecting the manner in which the early universe electroweak phase transition takes place~\cite{vonHarling:2017yew,Bruggisser:2018mus,Bruggisser:2018mrt}, and possibly playing a crucial role during post-inflationary reheating~\cite{Amin:2019qrx}.  Additionally, the different tunings have implications for the mass spectrum of the lightest scalar modes, and for general considerations about vacuum energy~\cite{Pomarol:2019aae,Cheung:2018xnu}.  For example, the radion mass may be typically lighter or heavier than the Higgs for the different classes of model.


\subsection{Higgs on the Brane}
We now want to consider the case in which the interactions between the Higgs and the GW scalar take place on the IR brane.  We assume no bulk interactions between the fields.  This can be done with a Higgs which is either in the bulk or completely localized on the IR brane.   We first consider the latter.

When the Higgs is localized on the IR brane, the only field that propagates in the bulk is the Goldberger-Wise (GW) scalar. The bulk action is then
\begin{equation}
S_\textnormal{bulk}=\int\dd[5]{x}\sqrt{g}\,\left(\frac{1}{2}g^{MN}\partial_M\Phi\,\partial_N\Phi+\frac{6}{\kappa^2}-\frac{1}{2} m_\Phi^2 \Phi^2-\frac{1}{2\kappa^2}R\right).
\end{equation}
We take $m_\Phi^2 \equiv \epsilon(\epsilon-4)$, where $\epsilon$ is taken to be ${\mathcal O}(1/10)$ in order to generate exponential hierarchies without severe tuning.  The bulk action is supplemented by brane-localized Lagrangian terms, including a kinetic term and potential for the Higgs on the IR brane:
\begin{equation}
S_\textnormal{brane}=-\int\dd[4]{x}\sqrt{-g_0}\,V_0(\Phi)\eval_{y=y_0}-\int\dd[4]{x}\sqrt{-g_1}\, \left(|\partial_\mu H |^2 - V_1(\Phi, |H|) \right)\eval_{y=y_1}.
\end{equation}
We presume the following form for the brane-localized potentials:
\begin{equation}
\begin{gathered}
V_0 = T_0 + \gamma_0 (\Phi-v_0)^2, \\
V_1 = T_1 + \lambda_H |H|^2 (|H|^2-v_H^2-\lambda \Phi) + \gamma_1 (\Phi-v_1)^2.
\end{gathered}
\end{equation}
These include brane tensions and localized mass terms for the Goldberger-Wise stabilizing field that displace the VEV from the origin.  In addition, there is a standard potential for the IR brane-localized Higgs with an additional trilinear coupling to the Goldberger-Wise field.  This last term is crucial for our purposes, as it couples the Higgs field to the radius.

We then solve the equation of motion~\eqref{genbulkeom}, supplemented by the boundary conditions~\eqref{genbcs}, under the assumption of an $x$-independent background VEV, $\langle \Phi \rangle = \phi(y)$.  Putting everything together we get the effective dilaton potential of Equation~\eqref{geneffpot} in the case of a single bulk scalar field, with the added contributions of the brane-localized Higgs.  

We consider the small backreaction limit, $\kappa \ll 1$. In this case, $A'\approx 1$ and the bulk equation of motion can be solved in general to give
\begin{equation}
\label{gw-sol}
\phi(y)=\phi_\epsilon e^{\epsilon y} +\phi_4 e^{(4-\epsilon)y}.
\end{equation}
The coefficients are fixed by imposing the boundary conditions. We adopt a stiff wall boundary condition for the GW field on the UV brane, corresponding to the limit $\gamma_0 \rightarrow \infty$, which fixes $\phi_0 \equiv \phi (y =0) = v_0$.

The IR brane potential contains the interaction between the GW scalar and the Higgs.  We do not take the stiff wall limit for the GW scalar on the IR brane, so $\phi_1 \equiv \phi(y_1)$ is not fixed at $v_1$.  Through the trilinear coupling to $\Phi$, the brane-localized Higgs mass is then a function of $y_1$.
The equation of motion for the Higgs background that minimizes the effective potential energy is
\begin{equation}
2 \langle H \rangle^2 \equiv v^2(y_1) = 
\begin{cases}
0 & v_H^2 + \lambda \phi_1 < 0 \\
v_H^2 + \lambda \phi_1 & v_H^2 + \lambda \phi_1 > 0,
\end{cases}
\label{eq:higgsvev}
\end{equation}
while the boundary condition for $\phi(y)$ is
\begin{equation}
\phi_1' = - \gamma_1 (\phi_1-v_1) + \frac{1}{4} \lambda \lambda_H v^2(y_1).
\end{equation}
If we impose only the IR boundary condition for $\phi$, leaving the Higgs VEV free (not imposing Equation~\eqref{eq:higgsvev}), we can explore the two dimensional effective potential as a function of $v$ and $y_1$.  

\begin{figure}
\centering
	\subfloat{\includegraphics[width=0.49\textwidth,keepaspectratio]{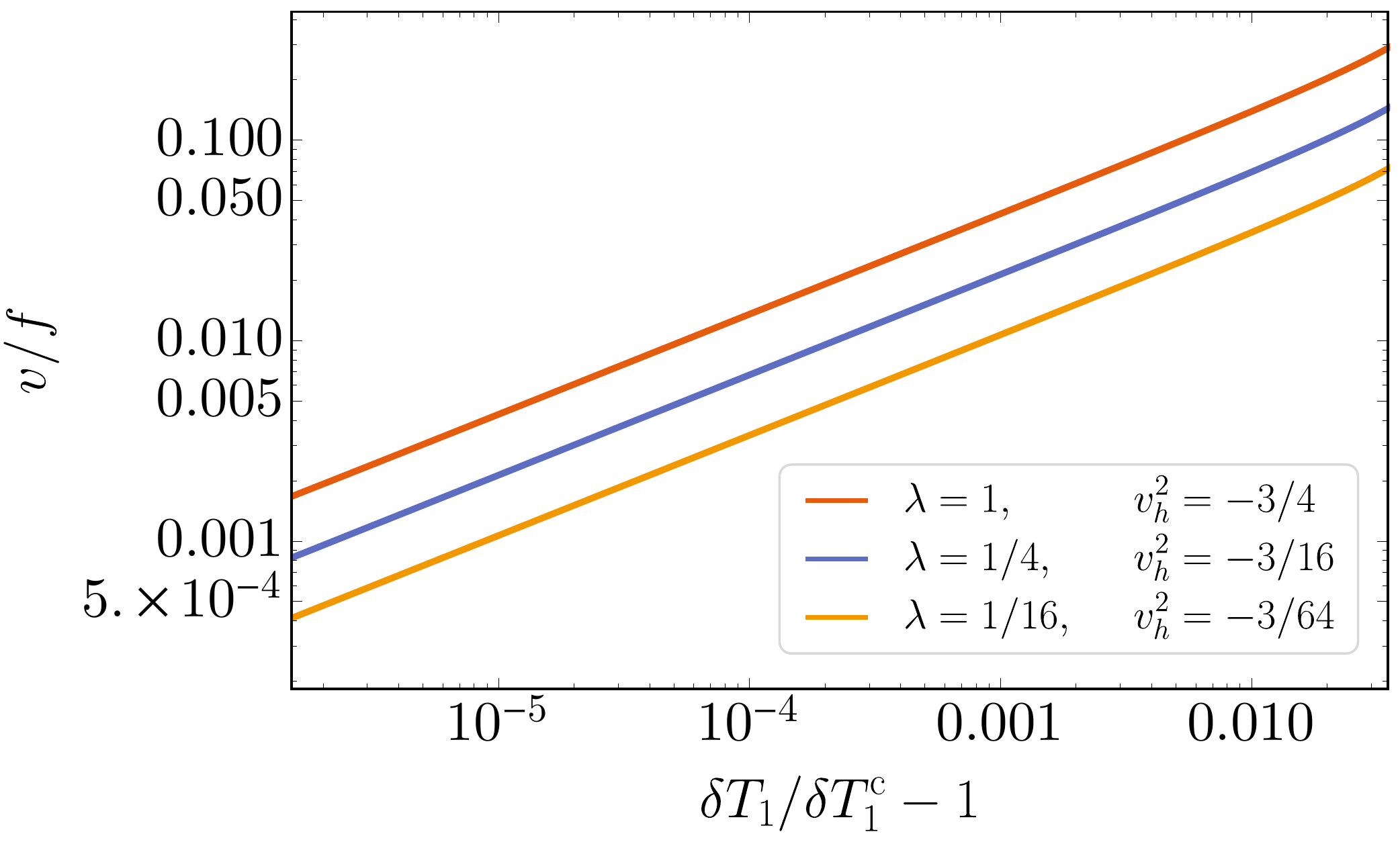}}\hfill%
	\subfloat{\includegraphics[width=0.49\textwidth,keepaspectratio]{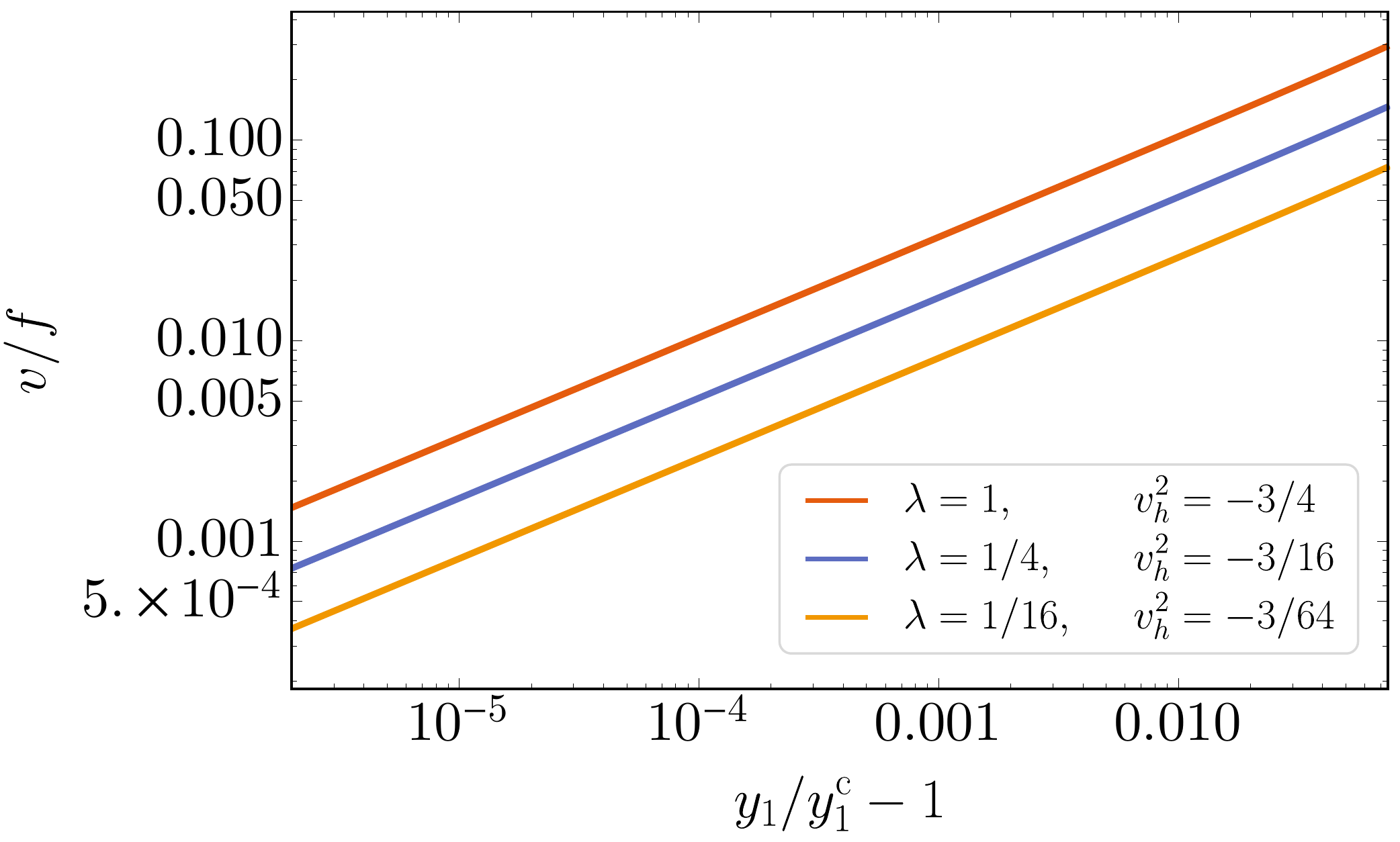}}
\caption{Higgs VEV in units of the symmetry breaking scale $f$ and its dependence on various parameters of the model. On the left, we show the variation as a function of the IR brane mistune $\delta T_1$. On the right, we display it in terms of how close the minimum of the radion potential is to $y_1^\textnormal{c}$, the critical point in the extra dimension where the interplay between the GW scalar and the Higgs first triggers electroweak symmetry breaking. We have taken $v_0=1/10$, $v_1=2$, $\gamma_1=1$, $\lambda_H=1/8$, $\kappa=1/50$. The different curves are obtained by fixing the ratio $\lambda/v_H^2$, so that equal values of $\epsilon$ correspond to equal values of $y_1^\textnormal{c}$.}
\label{fig:voverf}
\end{figure}

In the small backreaction limit, the effective potential for the system is
\beq
V_\textnormal{eff}=e^{-4y_0} \left( V_0 - \frac{6}{\kappa^2}-\frac{1}{4} {\phi'_0}^2- \frac{\epsilon (4-\epsilon)}{4}\phi_0^2\right)+e^{-4y_1}\left(V_1+\frac{6}{\kappa^2}+\frac{1}{4} {\phi'_1}^2+\frac{\epsilon(4-\epsilon)}{4}\phi_1^2\right).
\label{gw_eff_potential}
\eeq
For small $y_1$ values, it can be arranged that $v_H^2 + \lambda \phi_1 <0$, and the origin in Higgs field space is a stable minimum: $\expval{H}=0$. 

In this region, the effective potential (after imposing the Higgs equation of motion) is determined by the single GW field and its backreaction onto the geometry.

On the other hand, for larger $y_1$ values,  the Higgs may develop a VEV. 
For example, when $\epsilon$ is taken to be small and we set $\gamma_1=0$, we can approximate the electroweak symmetry breaking condition as
\begin{equation}
y_1 \gtrsim \frac{1}{\epsilon}\log(-\frac{v_H^2}{\lambda v_0}).
\end{equation}
Minimizing the effective potential with respect to $y_1$ relates the Higgs VEV to the symmetry breaking scale. The effective VEV, in units of the Kaluza-Klein scale $f$, scales with $\delta T_1$, the IR brane mistune, as
\begin{equation}
\label{vftune}
\frac{v}{f}\approx\left(\frac{64\delta T_1}{\lambda_H\left(16-\lambda^2\lambda_H\right)}\right)^{1/4}.
\end{equation}

In Figure~\ref{fig:voverf} we quantify the amount of tuning associated with a given Higgs VEV $v$. The Higgs VEV is fixed in the following way: first we identify the effective 4D scale of gravity, i.e.\ the Planck scale, via $M_\textnormal{Pl}^2 \approx 1/(2\kappa^2)$, as a consequence of the warped geometry. For a given value of $\kappa$, we then reproduce the experimentally measured hierarchy between the Planck and the weak scale, $\log(M_\textnormal{Pl}/v)\approx38.4$. We scan over $f$ by changing $\epsilon \sim 1/10$, so that the corresponding values of $y_1$ provide a large range of Kaluza-Klein scales.  To obtain a small $v/f$, parameters must be chosen so that the global minimum of the effective potential is very close to the critical point for electroweak symmetry breaking.  We denote the critical values of the 5D parameters with the superscript ``c".  We display this information in two different ways: on the left, as a function of $\delta T_1$, which is the physical parameter that needs to be tuned to adjust the location of the minimum of the effective potential. On the right, as a function of $y_1$, the location of the minimum itself, which is more useful to visually determine how close the minimum has to be to the critical point. As expected from the analytic estimate of Equation~\eqref{vftune}, a small Higgs VEV requires a very small mistune in the brane tension against the bulk cosmological constant.


\subsubsection*{Mass Spectrum}
\label{mass-spectrum-higgs-on-the-brane}
In order to understand the spectrum of the theory, we consider fluctuations around the background solutions for the metric and the two scalar field profiles to linearized order. In particular, we parametrize the various degrees of freedom as
\begin{equation}
\begin{gathered}
\label{fluct}
\Phi(x,y)=\phi(y)+\varphi(x,y),   \\
H(x)=\frac{1}{\sqrt{2}}\left[v+h(x)\right]\exp(i\alpha),   \\
\dd{s^2}=e^{-2A(y)-2F(x,y)}\eta_{\mu\nu}\dd{x^\mu}\dd{x^\nu}-(1+2F(x,y))^2\dd{y^2}.
\end{gathered}
\end{equation}
It is possible to show that, when having only one scalar field in the bulk, the whole set of linearized Einstein equations, together with the equations of motion for the scalar fields, reduces to a single homogeneous differential equation for $F$ in the bulk~\cite{Csaki:2000zn}:
\begin{equation}
\label{eigenbulk}
F''-2A'F'-4A''F-2\frac{\phi''}{\phi'}F'+4\frac{\phi''}{\phi'}A'F=e^{2A}\Box F.
\end{equation}
We also have to specify the boundary conditions. When considering a stiff wall UV brane boundary condition for the background GW field ($\gamma_0 \rightarrow \infty$), the UV boundary condition is $F_0'=2A_0'F_0$.  The IR boundary condition for $F$ can be obtained by combining the following equations:
\begin{equation}
\begin{gathered}
\label{IRBC}
\phi'\varphi=\frac{3}{\kappa^2}\left(F'-2A'F\right),   \\
-2\varphi_1'=\pdv[2]{V_1}{\phi}\varphi_1+\pdv{V_1}{\phi}{h}h+2F_1\pdv{V_1}{\phi},   \\
\Box h+e^{-2A_1}\left(\pdv{V_1}{\phi}{h}\varphi_1+\pdv[2]{V_1}{h}h\right)=0.
\end{gathered}
\end{equation}
Notice that the mass eigenvalue appears both in the bulk equation and in the boundary condition because of the presence of the brane-localized Higgs kinetic term which leads to the $\Box\equiv\partial_\mu\partial^\mu$ operator.

\begin{figure}
\centering
	\subfloat{\includegraphics[width=0.49\textwidth,keepaspectratio]{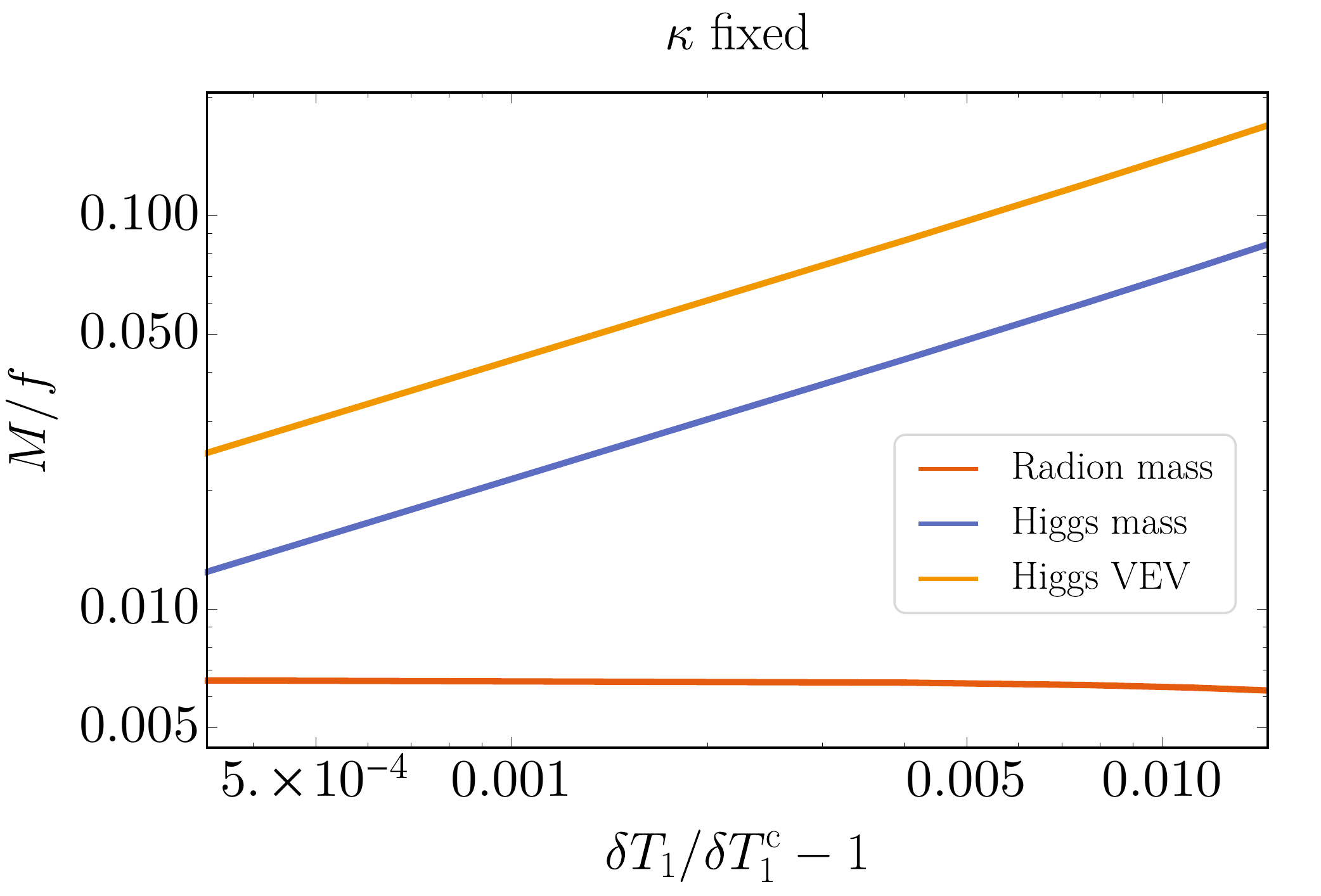}}\hfill%
	\subfloat{\includegraphics[width=0.49\textwidth,keepaspectratio]{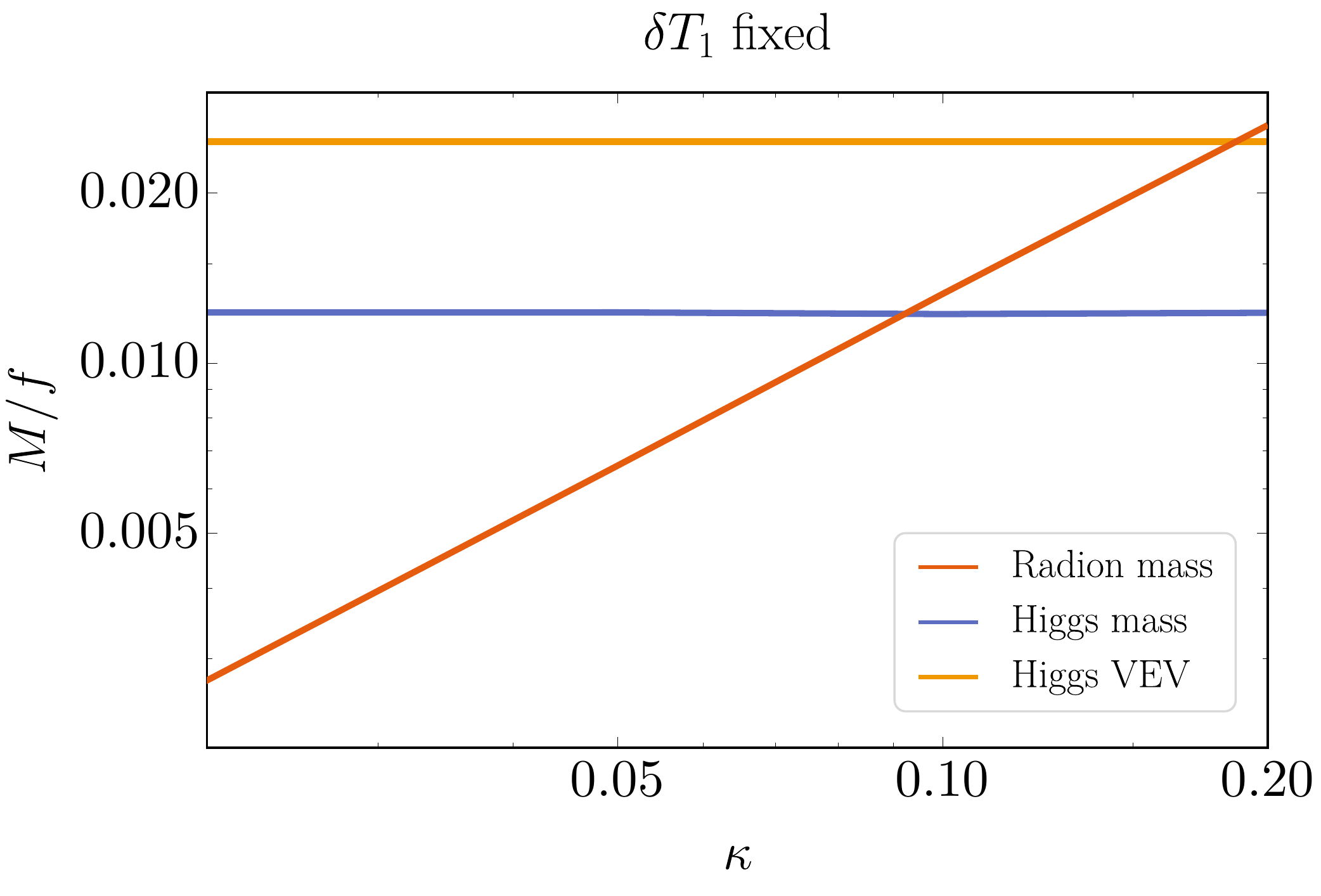}}
\caption{Dependence of the mass scales in the theory on various parameters of the model. On the left, we show how for a fixed value of $\kappa$ the IR brane mistune controls the Higgs mass and VEV, while the radion mass is essentially unchanged. On the right, we see that the radion mass is dependent on the amount of backreaction. The parameters are the same as in Figure~\ref{fig:voverf}.}
\label{fig:eigenbrane}
\end{figure}

Solutions are parametrized by 3 unknowns.   Two are integration constants associated with the second order equation for $F$, and the third is the eigenvalue.  One integration constant can be fixed by choosing a convenient normalization, and the other is eliminated by imposing the UV brane boundary condition.  To find the eigenvalues, we employ a shooting method and determine the mass eigenvalues for which the IR boundary condition is satisfied.  We find that the spectrum contains a light radion and Higgs along with a KK tower due to the bulk GW field.\footnote{As usual, there is no separate KK tower for the radion field, as these modes are eaten by the massive KK gravitons.}\footnote{Recent work also utilizes a new approach to get the full potential for the lightest scalar mode, going beyond the mass eigenvalue~\cite{Lizana:2019ath}.}

Figure~\ref{fig:eigenbrane} shows the mass spectrum obtained by solving Equation~\eqref{eigenbulk} supplemented by the boundary conditions~\eqref{IRBC}. Once again, we fix $v$ to the electroweak value and we scan over $f$ by varying $\epsilon\sim1/10$. We can see how the mass of the Higgs fluctuation tracks the Higgs VEV and is controlled by the amount of mistune (left subfigure), while the radion mass is determined by the backreaction parameter $\kappa$ (right subfigure). On the other hand, the Higgs and radion masses are essentially insensitive to changes in $\kappa$ and $\delta T_1$, respectively.  The lightest particle in the spectrum can be either the Higgs or the radion, depending on the choice of parameters.  This is in contrast to the results of Subsection~\ref{sec:higgs_only_in_the_bulk}, where we consider only the Higgs in the bulk, and the radion is constrained to be the lightest state in the spectrum.

\subsection{Two Bulk Fields}
In this Subsection, we consider the case where both fields, GW scalar $\Phi$ and the Higgs $H$, propagate in the bulk. For simplicity, we assume that they are uncoupled, despite such a coupling being allowed by the symmetry. However, they still interact indirectly through gravity, as both fields backreact on the geometry. We assume that the Higgs is localized towards the IR brane by taking its bulk mass close to (but above) the Breitenlohner-Freedman bound $m_H^2 = -4$. Note that the bulk mass is constant. 

The bulk action is given by
\begin{align}
	S_{\text{bulk}}=\int \dd[5]{x}\sqrt{g}\left(\frac{1}{2}\partial^N \Phi \partial_N \Phi + \partial^N H^{\dagger}\partial_N H + \frac{6}{\kappa^2}-\frac{1}{2}m_{\Phi}^2\Phi^2-m_H^2 \abs{H}^2-\frac{1}{2 \kappa^2}R\right).
\end{align}
Again we take $m_{\Phi}^2 \equiv \epsilon(\epsilon-4)$ with $\epsilon \sim \mathcal{O}(1/10)$, and we also define $\nu^2 \equiv 4 + m_H^2$ with $\nu \sim \mathcal{O}(1/10)$. This bulk action is supplemented by brane-localized Lagrangian terms
\begin{align}
	S_{\text{brane}}=-\int \dd[4]{x}\sqrt{-g_0}V_0(\Phi,\abs{H})\eval_{y=y_0}-\int \dd[4]{x}\sqrt{-g_1}V_1(\Phi,\abs{H})\eval_{y=y_1}.
\end{align}
The brane-localized potentials are given by the following:
\beq
\begin{gathered}
V_0(\Phi,\abs{H})=T_0 + \gamma_0 (\Phi - v_0)^2 + m_0^2 \abs{H}^2, \\
V_1(\Phi,\abs{H})=T_1 + \gamma_1 (\Phi - v_1)^2 + \lambda_H \abs{H}^2 (\abs{H}^2 - v_H^2).
\label{brane_potentials}
\end{gathered}
\eeq
In this Subsection, we work in the simplifying limit $\gamma_{0,1} \rightarrow \infty$, which sets the UV/IR boundary conditions for the GW field as $\Phi(y=y_0)=v_0$ and $\Phi(y=y_1)=v_1$ respectively. 

We write the VEVs for the GW and the Higgs field respectively as $\expval{\Phi}=\phi(y)$ and $\expval{\abs{H}}=v(y)/\sqrt{2}$. Presuming small metric backreaction, the bulk equations of motion for the scalars are given by
\beq
\begin{gathered}
	\phi''-4\phi'-\epsilon(\epsilon-4)\phi=0, \\
	v''-4v'-(-4+\nu^2)v=0.
\end{gathered}
\eeq
The GW field $\phi$ has the same profile as in the previous Subsection, which is given by \eqref{gw-sol} where the coefficients $\phi_{\epsilon}$ and $\phi_4$ are fixed by the boundary conditions $\phi(y_0= 0)=v_0$ and $\phi(y_1)=v_1$. The solution for the Higgs VEV can be conveniently expressed as
\beq
	\label{higgs_sol}
	v(y)=v(y_1)e^{2(y-y_1)} \left(\frac{e^{\nu y} - r e^{-\nu y}}{e^{\nu y_1} - r e^{-\nu y_1}}\right),
\eeq
where $v(y_1)$ and $r$  are integration constants which will be determined by the Higgs boundary conditions generated by the brane potentials \eqref{brane_potentials}:
\beq
\begin{gathered}
	v'(y_0)=\frac{m_0^2}{2}v(y_0), \\
	v'(y_1)=\frac{\lambda_H}{2}v(y_1)\left(v_H^2-v^2(y_1)\right).
\end{gathered}
\eeq
The UV boundary condition fixes $r$ as
\begin{align}
	\label{fix_r}
	r=\frac{m_0^2-4-2\nu}{m_0^2-4+2\nu}.
\end{align}
The constant $v(y_1)$, the Higgs VEV on the IR brane, is determined by the IR boundary condition which has two solutions. One of them is the trivial one $v(y_1)=0$ corresponding to unbroken electroweak symmetry. By defining $m_1^2 \equiv \lambda_H v_H^2$, the second solution is given by 
\beq
	\label{higgs_vev_on_ir}
	v^2(y_1)=\frac{1}{\lambda_H}\left[(m_1^2-4-2\nu)+\frac{4\nu (m_0^2-4-2\nu)}{(m_0^2-4-2\nu)-e^{2\nu y_1}(m_0^2-4+2\nu)}\right],
\eeq
provided that the term inside brackets is positive. In this case, the above solution is the preferred one, corresponding to broken electroweak symmetry. 

If the Higgs has a nonzero VEV, it contributes to the 4D effective potential \eqref{gw_eff_potential} by a term given by
\beq
	V_{\text{eff}}^v=\left[\frac{m_0^2}{2}v^2-\frac{1}{4}v'^2+\frac{1}{4}m_H^2 v^2\right]\eval_{y=0}+e^{-4 y_1}\left[ \frac{\lambda_H}{2}v^2 \left(\frac{v^2}{2} - v_H^2\right)+\frac{1}{4}v'^2-\frac{1}{4}m_H^2 v^2\right]\eval_{y=y_1}
\eeq
By using the Higgs solution \eqref{higgs_sol} and the boundary conditions \eqref{fix_r} and \eqref{higgs_vev_on_ir}, the above expression takes a very simple form:
\beq
	\label{higgs_eff_pot_final}
	V_{\text{eff}}^v=-\frac{\lambda_H}{4}v^4(y_1)e^{-4 y_1}.
\eeq
We can directly see that it is negative definite, therefore if there is a solution with a non-trivial VEV, then it will be energetically favored.  

\subsubsection*{Parameter Space for Electroweak Symmetry Breaking}
\label{sec:ew_parameter_space_gw_higgs}

By inspecting the function $v^2(y_1)$, we can see that it has a singularity at
\begin{align}
	y_1^s=\frac{1}{2\nu }\log \left(\frac{m_0-4-2\nu}{m_0^2-4+2\nu}\right),
\end{align}
since the denominator of the second term in \eqref{higgs_vev_on_ir} diverges. In the region of parameter space where $m_0^2-4<-2\nu$, $y_1^s$ is positive, hence physical. As a result, the effective potential will be unbounded from below at $y_1=y_1^s$. This singularity is an artifact of neglecting the backreaction of the Higgs on the geometry, which cannot be done in the vicinity of $y_1^s$. Nevertheless, since we want to continue to work in the small backreaction limit, we will exclude the $m_0^2-4<-2\nu$ from our parameter space. 

For the rest of the discussion, we will assume $m_0^2-4>-2\nu$ and define $\alpha_{0,1}\equiv m_{0,1}^2-4-2\nu$ for notational simplicity. By taking the derivative of \eqref{higgs_vev_on_ir} with respect to $y_1$, we find
\begin{align}
\label{vev_derivative}
	\pdv{v^2(y_1)}{y_1}=\frac{1}{\lambda_H}\frac{8 \alpha_0 \nu^2(\alpha_0 + 4\nu)e^{2 \nu y_1}}{[\alpha_0 - e^{2\nu y_1}(\alpha_0+4\nu)]^2}.
\end{align}
This tells us that $v^2(y_1)$ is either monotonically increasing or decreasing depending on the sign of $\alpha_0$. We will consider the former case in this particular example.

In order to have symmetry breaking, we need $\lim_{y_1 \rightarrow \infty} v^2(y_1)>0$ which implies $\alpha_1>0$. Additionally, the symmetry will be unbroken in the UV if $v^2(y_1=0)<0$, or $\alpha_0>\alpha_1$. In this case there is a ``critical" position of the IR brane, $y_1=y_1^c$, where the effective Higgs mass squared term is vanishing. This point corresponds to $v^2(y_1=y_1^c)=0$ and is given by
\begin{align}
	y_1^c=\frac{1}{2\nu}\log \left(\frac{\alpha_0 (\alpha_1 + 4\nu)}{\alpha_1 (\alpha_0+4\nu)}\right).
\end{align}

A convenient measure of the size of the Higgs VEV is the mass that would be given to a gauge field by the Higgs mechanism. For small backreaction, this can be approximated by
\begin{align}
	\label{eff_higgs_vev}
	\left(\frac{v_{\text{eff}}}{f}\right)^2=\int_{y_0}^{y_1}\dd{y}e^{-2(y-y_1)}v(y)^2,
\end{align}
where $f=e^{-y_1}$ is the conformal symmetry breaking scale, i.e. the KK scale.

The full effective potential can be expressed as the sum of the UV and IR contributions:
\begin{align}
	V_{\text{eff}}(y_1)=e^{-4 y_0}V_{\text{eff}}^{\text{UV}}(y_1)+e^{-4y_1}V_{\text{eff}}^{\text{IR}}(y_1).
\end{align}
As we have proved in Appendix~\ref{sec:derivativederivation}, the condition for an extremum of the effective potential is that $V_{\text{eff}}^{\text{IR}}$ vanishes, provided that the scalar boundary conditions \eqref{genbcs} are satisfied. Such a point, let us denote it by $y_1^{\text{ext}}$, is a minimum provided that
\begin{align}
	V_{\text{eff}}''(y_1)\eval_{y_1=y_1^{\text{ext}}}>0.
\end{align}
Since the Higgs contribution to the effective potential scales as the fourth power of the Higgs VEV, we can neglect it for finding the extremum points and their stability. Then we can approximate $V_{\text{eff}}^{\text{IR}}$ by
\begin{align}
	V_{\text{eff}}^{\text{IR}}(y_1)\approx \delta \tilde{T}_1 - v_0 v_1 (2-\epsilon)(4-\epsilon)e^{\epsilon y_1}+v_0^2(2-\epsilon)^2 e^{2 \epsilon y_1},
\end{align}
where in the second line we absorbed all the $y_1$-independent terms into the definition of $\delta \tilde{T}_1$. The extremum of the potential is determined by the solution of the quadratic equation:
\begin{align}
	\label{gw_extremum}
	\delta \tilde{T}_1 - v_0 v_1 (2-\epsilon)(4-\epsilon)f_{\text{ext}}^{-\epsilon}+v_0^2(2-\epsilon)^2 f_{\text{ext}}^{-2\epsilon}=0,
\end{align}
where $f_{\text{ext}}= \exp{-y_1^{\text{ext}}}$ denotes the conformal breaking scale at the extremum point $y_1^{\text{ext}}$. To determine the stability, we need to calculate the full effective potential. Again ignoring Higgs contributions, it can approximately be expressed as
\begin{align}
	V_{\text{eff}}(f)\approx \delta \tilde{T}_0 + \delta \tilde{T}_1 f^4 - 4 v_0 v_1 (2-\epsilon)f^{4-\epsilon}+2 v_0^2(2-\epsilon)f^{4-2\epsilon}.
\end{align}
Since $V_{\text{eff}}''(y_1)$ and $V_{\text{eff}}''(f)$ have the same sign, we can calculate the latter. Then, the stability condition reads
\begin{align}
	12 \delta \tilde{T}_1 - 4 v_0 v_1 (2-\epsilon)(3-\epsilon)(4-\epsilon)f^{-\epsilon}_{\text{ext}}+2 v_0^2(2-\epsilon)(3-2\epsilon)(4-2\epsilon)f^{-2\epsilon}_{\text{ext}}>0.
\end{align}
Finally, we replace $\delta \tilde{T}_1$ with the solution of \eqref{gw_extremum}. Then we find that the extremum point $f_{\text{ext}}$ is stable if
\begin{align}
	\label{stability_condition}
	f_{\text{ext}}<\left(\frac{(v_1/v_0)(4-\epsilon)}{4-2\epsilon}\right)^{-1/\epsilon}.
\end{align}

\subsubsection*{Mass Spectrum}
In this Subsection, we will work out the mass spectrum when both fields are propagating in the bulk. We will parameterize the fluctuations of the metric and the field in the same way as in Subsection~\ref{mass-spectrum-higgs-on-the-brane}, except now the Higgs field does also depend on the bulk coordinate $y$:
\begin{align}
	H(x,y)=\frac{1}{\sqrt{2}} \left[v(y)+h(x,y)\right]\exp{i\alpha}.
\end{align}
In this case, the linearized Einstein equations for fields $F, h, \varphi$ take the form
\begin{align}
	\label{eq:6}
	12 A'^2F-6 A''F-6A'F'+F''&=-\frac{\kappa^2}{3}\left[ 2 VF+3F(\phi'^2+v'^2)-\pdv{V}{\phi}\varphi - \pdv{V}{v}h-\phi'\varphi'-v'h' \right]\nonumber \\
	&\phantom{{}={}}-\frac{\kappa^2}{3}\sum_{j=0,1}\left( 4 V_j F - \pdv{V_j}{\phi}\varphi - \pdv{V_j}{v}h\right)\delta(y-y_j),\\
	\label{eq:7}
	4 A' F'+ e^{2A}\square F &= -\frac{\kappa^2}{3} \left(4 VF+\pdv{V}{\phi}\varphi + \pdv{V}{v}h - \phi'\varphi'-v'h'  \right),\\
	\label{eq:8}
	\partial_{\mu}\left( 3 F'-6A'F \right)&=\kappa^2 \partial_{\mu}\left( \phi'\varphi+v'h \right),
\end{align}
where all derivatives of the bulk and brane potentials are evaluated on the background scalar VEVs. The $\mu5$-equation can directly be integrated to give
\begin{align}
	\label{eq:9}
	F'-2 A'F=\frac{\kappa^2}{3}\left( \phi'\varphi+v'h \right).
\end{align}
By combining the $\mu\nu$- and $55$-equations in the bulk we obtain 
\begin{align}
	\label{eq:10}
	F''-2A'F'+e^{2A}\square F=\frac{2\kappa^2}{3}\left( \phi'\varphi'+v'h' \right).
\end{align}
Matching the singular terms in the $\mu\nu$-equation gives the junction conditions for $F$:
\begin{align}
	\label{eq:11}
	[F']_i=\frac{2\kappa^2}{3}V_i F+\frac{\kappa^2}{3} \left(\pdv{V_i}{\phi}\varphi+\pdv{V_i}{v}h  \right).
\end{align}
By using the boundary conditions for the background solution, one can show that this is equivalent to the $\mu5$-equation so it provides no new constraints.

The linearized scalar field equations are given by
\begin{gather}
	\label{eq:12}
	e^{2A}\square \varphi - \varphi''+4A'\varphi'+\pdv[2]{V}{\phi} \varphi=-6F'\phi'-4\pdv{V}{\phi}F,\\
	\label{eq:13}
	e^{2A}\square h - h''+4A' h'+\pdv[2]{V}{v} h=-6F'v'-4\pdv{V}{v}F,
\end{gather}
together with the boundary conditions
\begin{align}
	\label{eq:14}
	[\varphi' -2\phi' F]_i&=\pdv[2]{V_i}{\phi}\eval_{\phi}\varphi,\\
	\label{eq:15}
	[h'-2v'F]_i&=\pdv[2]{V_i}{v}h.
\end{align}
So far, there are three second order differential equations, (\ref{eq:10})(\ref{eq:12}) and (\ref{eq:13}), which need to be solved simultaneously. However, we can use the $\mu5$-equation (\ref{eq:9}) to eliminate $\varphi$ from the system. Then (\ref{eq:10}) becomes
\begin{align}
\label{eq:17}
F''-2A'F'-4A''F-2 \frac{\phi''}{\phi'}F'+4A' \frac{\phi''}{\phi'}F=e^{2A}\square F + \frac{2\kappa^2}{3}\left( v''-\frac{\phi''}{\phi'}v' \right)h.
\end{align}

To find the mass spectrum, we expand both $F$ and $h$ into their Kaluza-Klein (KK) modes by
\begin{align}
	F(x,y)&=\sum_n F_n(y)R_n(x),\\
	h(x,y)&=\sum_n h_n(y)R_n(x),
\end{align}
where each KK mode in the above expansions satisfies $\square R_n=-m_n^2 R_n$. Using \eqref{eq:13} and \eqref{eq:17}, we write the system of differential equations to determine the mass spectrum as
\begin{gather}
	\label{mass-spectrum-f}
	F_n''-2A'F_n'-4A''F_n-2 \frac{\phi''}{\phi'}F_n'+4A' \frac{\phi''}{\phi'}F_n+e^{2A}m_n^2 F_n = \frac{2\kappa^2}{3}\left( v''-\frac{\phi''}{\phi'}v' \right)h_n,\\
	\label{mass-spectrum-h}
	h_n''-4A' h_n'-\left(\pdv[2]{V}{v} -e^{2A}m_n^2 \right) h_n=6F_n'v'+4\pdv{V}{v}F_n.
\end{gather}
This system of differential equations have five integration constants; two from each equation plus the eigenvalue. Two of them are fixed by the boundary condition \eqref{eq:15}, which reads
\begin{align}
	h_n'-2v'F_n = \pm \frac{1}{2}\pdv[2]{V_{0,1}}{v}h_n,\quad \text{$+/-$ is for the UV/IR brane}.
\end{align}
In the case of stiff-wall boundary conditions, \eqref{eq:14} sets $\varphi=0$ on both branes. Then, \eqref{eq:11} implies the relation
\begin{align}
	F_n'-2A'F_n=\frac{\kappa^2}{3}v'h_n,\quad \text{on the branes}.
\end{align}
The remaining integration constant can be fixed by normalizing both $F_n$ and $h_n$ by a common factor, since the system is invariant under such a scaling.

The rest of this Subsection is devoted to solving this system of equations to zero and leading order in the backreaction. We will assume that $\kappa$ is small enough so that the background field profiles are accurately expressed by their zero backreaction solutions. We shall investigate the validity of this assumption later.


\subsubsection*{Mass Spectrum with Backreaction Neglected}
\label{sec:mass-spectrum-with}
First, we will study the mass spectrum in the $\kappa^2 \rightarrow 0$ limit. This means that we will search for the mass eigenvalues which remain finite after this limit. We will denote these eigenvalues and their corresponding eigenvector components $F_n$ and $h_n$ by the superscript $(0)$. Then \eqref{mass-spectrum-f} and \eqref{mass-spectrum-h} become
\begin{gather}
	\label{mass-spectrum-f-0}
	\left(F_n^{(0)'}-2F_n^{(0)}\right)'-2 \frac{\phi''}{\phi'}\left(F_n^{(0)'}-2 F_n^{(0)}\right)+e^{2y}\left(m_n^{(0)}\right)^2 F_n^{(0)} =0,\\
	\label{mass-spectrum-h-0}
	h_n^{(0)''}-4 h_n^{(0)'}-\left(\pdv[2]{V}{v} -e^{2y}\left(m_n^{(0)}\right)^2 \right) h_n^{(0)}=6F_n^{(0)'}v'+4\pdv{V}{v}F_n^{(0)}.
\end{gather}
The $\mu5$-component of the Einstein equations \eqref{eq:9} tells us that $F_n^{(0)'}-2F_n^{(0)} \propto \kappa^2$. Then \eqref{mass-spectrum-f-0} implies $F_n^{(0)} \propto \kappa^2$ too. Hence, the mass spectrum in the $\kappa^2 \rightarrow 0$ limit is given by the differential equation
\beq
	\label{eq:28}
	h_n''-4 h_n' +(m_n^2e^{2y}- m_H^2)h_n=0,
\eeq
with boundary conditions
\beq
	\label{eq:29}
	h'_n=\pm\frac{1}{2}\pdv[2]{V_{0,1}}{v}h_n,
\eeq
where $+/-$ for UV/IR brane, and we have omitted the superscripts for brevity.

We see that the fluctuations $F$ and $h$ are decoupled from each other in this limit, which is expected since there is no explicit coupling between the GW and the Higgs field. In the absence of backreaction, each field does not know about the existence of the other. 

The solution of \eqref{eq:28} is given in terms of Bessel functions:
\begin{align}
	\label{eq:30}
	h_n=e^{2y} \left[J_{\nu}(m_n e^y)+c Y_{\nu}(m_ne^y) \right],
\end{align}
where $c$ is a constant which is determined by the UV boundary condition. The mass eigenvalue is fixed by the IR boundary condition. After applying both, we find that the mass spectrum is given by the roots of
\begin{align}
	\label{eq:33}
	b_v(m_n)=\tilde{J}_{\nu}^1(m_n)-\frac{\tilde{J}_{\nu}^0(m_n)}{\tilde{Y}_{\nu}^0(m_n)}\tilde{Y}_{\nu}^1(m_n),
\end{align}
where we defined
\begin{gather}
	\tilde{X}^0(m_n)\equiv\left(\pdv[2]{V_0}{v}-4\right)X_{\nu}(m_n)-m_n \left(X_{\nu-1}(m_n)-X_{\nu+1}(m_n)\right),\\
	\tilde{X}^1(m_n)\equiv\left(\pdv[2]{V_1}{v}+4\right)X_{\nu}(m_n)+m_n e^{y_1} \left(X_{\nu-1}(m_n e^{y_1})-X_{\nu+1}(m_n e^{y_1})\right),
\end{gather}
with $X=\qty{J,Y}$. The Higgs mass is the smallest $m_n$ which satisfy $b_v(m_n)=0$.
\begin{figure}
	\centering
	\subfloat{\includegraphics[width=0.48\textwidth,keepaspectratio]{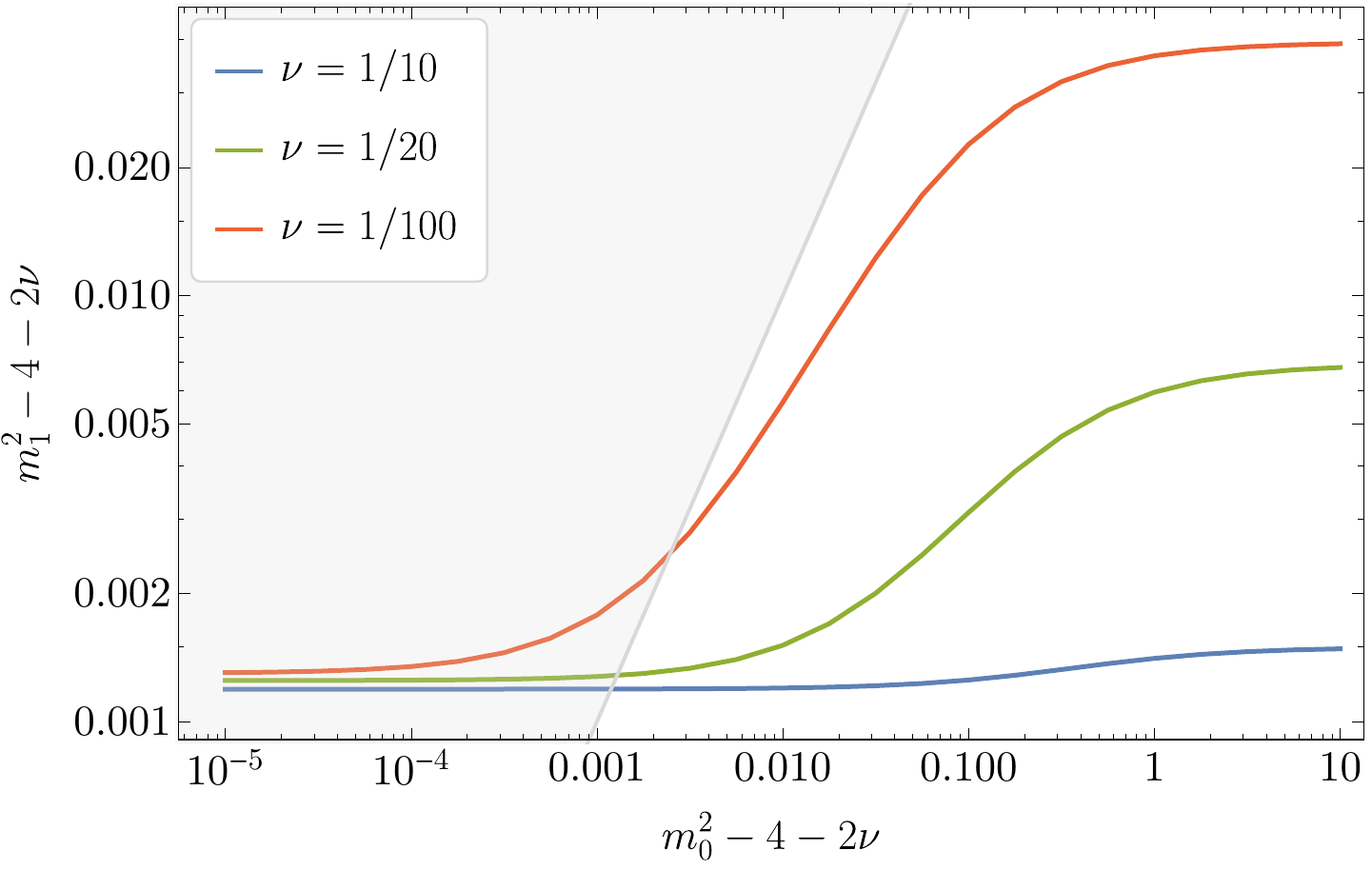}}\hfill%
	\subfloat{\includegraphics[width=0.48\textwidth,keepaspectratio]{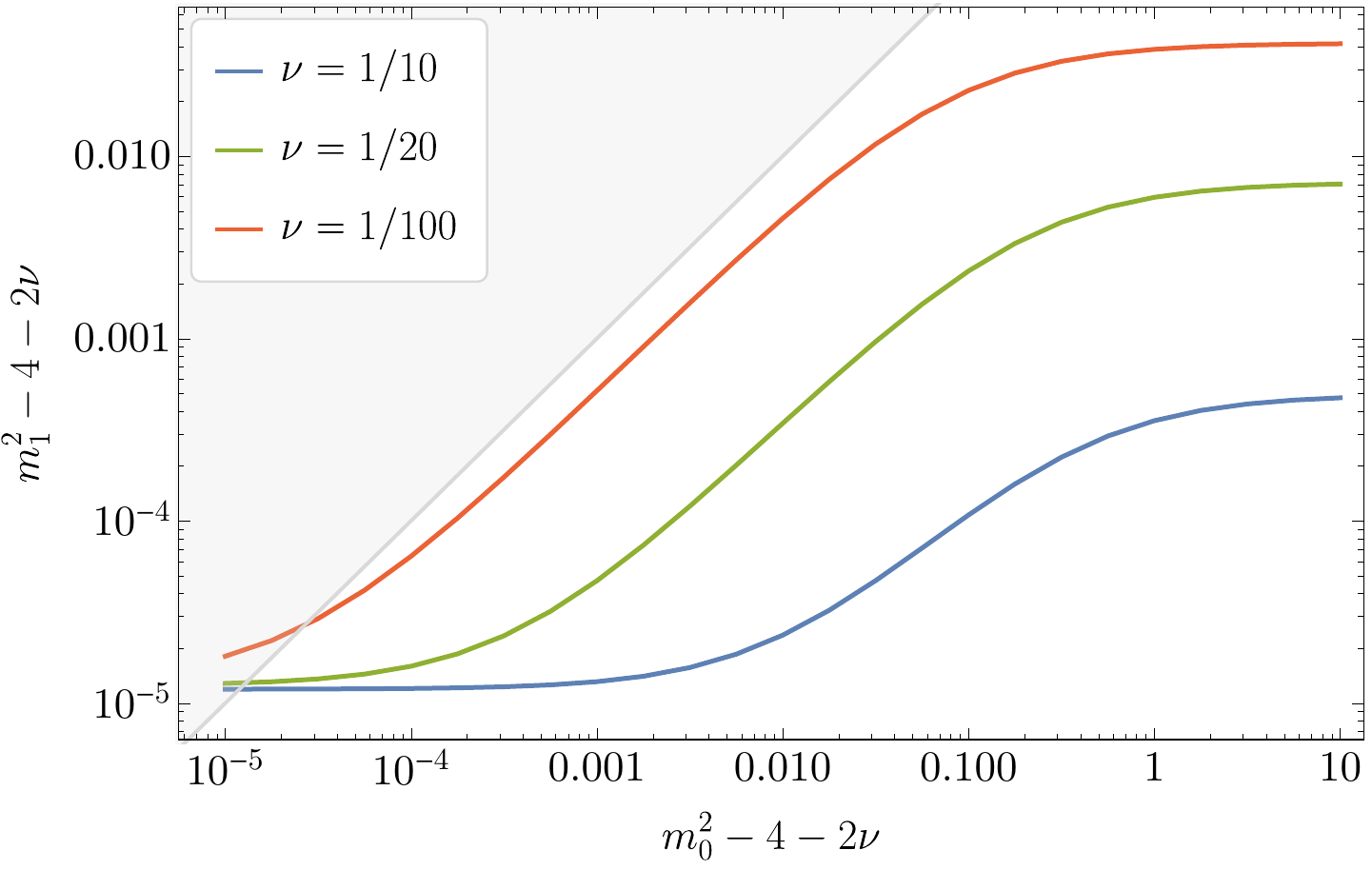}}
	\caption{In this figure, we are plotting the necessary value of $m_1^2$ in order to get $v_{\text{EW}}=246\;\si{\giga\electronvolt}$ and $m_h=126\;\si{\giga\electronvolt}$, as a function of $m_0^2$. The hierarchy between the electroweak and the conformal breaking scale is $f/v_{\text{EW}}=10$ and $f/v_{\text{EW}}=100$ for the left and right plot respectively. In the shaded region, there is no "critical region", i.e.\ the electroweak symmetry is broken, even if the IR brane is very close to the UV brane.}
	\label{fig:higgs-param-space}
\end{figure}

The Higgs sector of the model is specified by five parameters; $m_0^2$, $m_1^2$, $\lambda_H$, $\nu$ and $f$. Two of them can be fixed in terms of the others, by setting the effective Higgs VEV \eqref{eff_higgs_vev} and the Higgs mass to $246\;\si{\giga\electronvolt}$ and $126\;\si{\giga\electronvolt}$ respectively. We have chosen to keep $m_0^2$, $\nu$ and $f$ free, and calculate $m_1^2$ and $\lambda_H$ in terms of the rest. We show the results in Figure~\ref{fig:higgs-param-space}. One can see that a fair amount of tuning is needed in the IR brane mass parameter of the Higgs. One might get the impression that the tuning is less severe when we lower the $\nu$ parameter, or increase the hierarchy between the electroweak and conformal breaking scales. But in that case, the tuning between the GW and the Higgs sector does increase, as we shall see shortly. 

\begin{figure}
	\centering
	\subfloat{\includegraphics[width=0.48\textwidth,keepaspectratio,height=5cm]{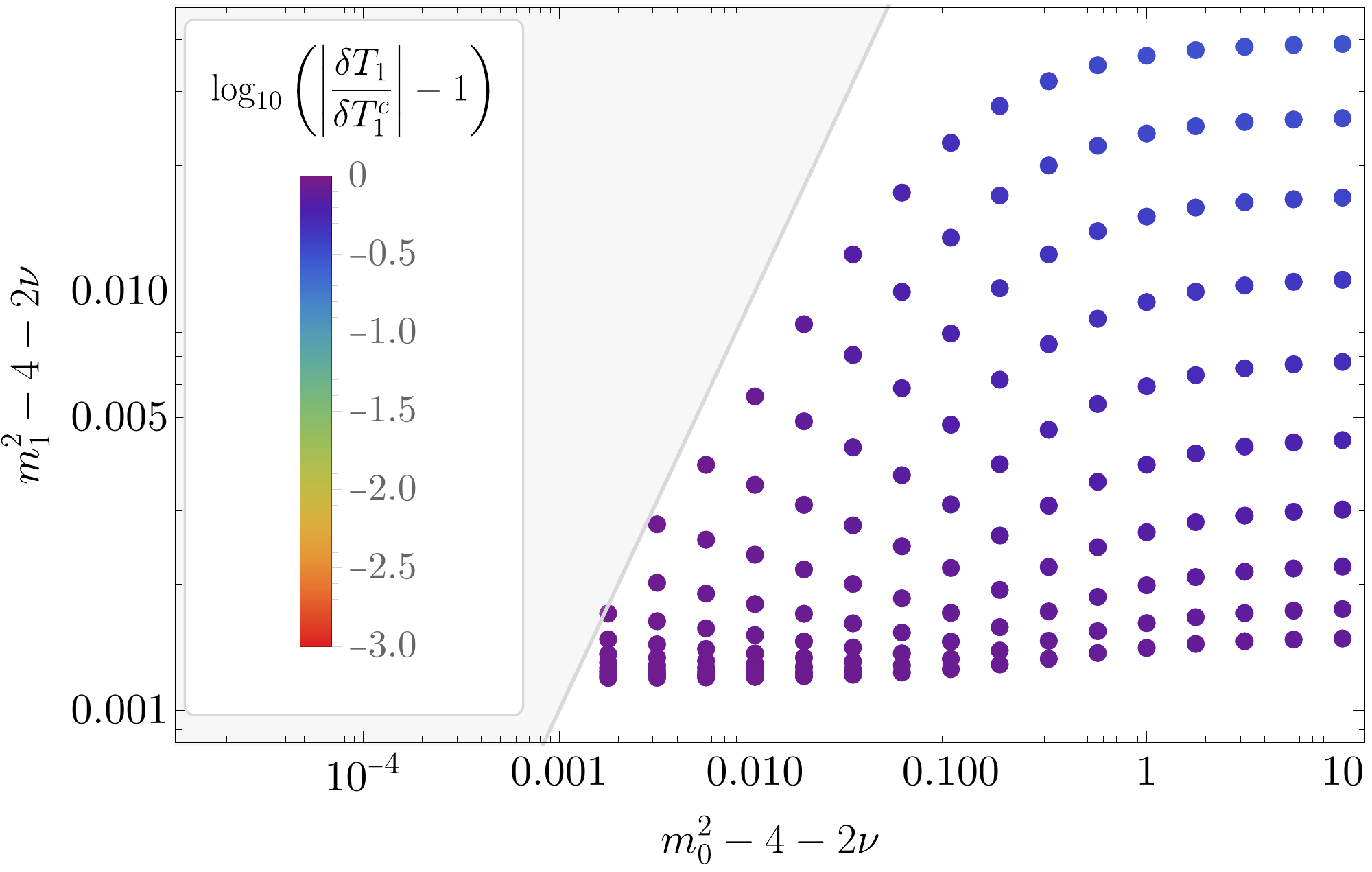}}\hfill%
	\subfloat{\includegraphics[width=0.48\textwidth,keepaspectratio,height=5cm]{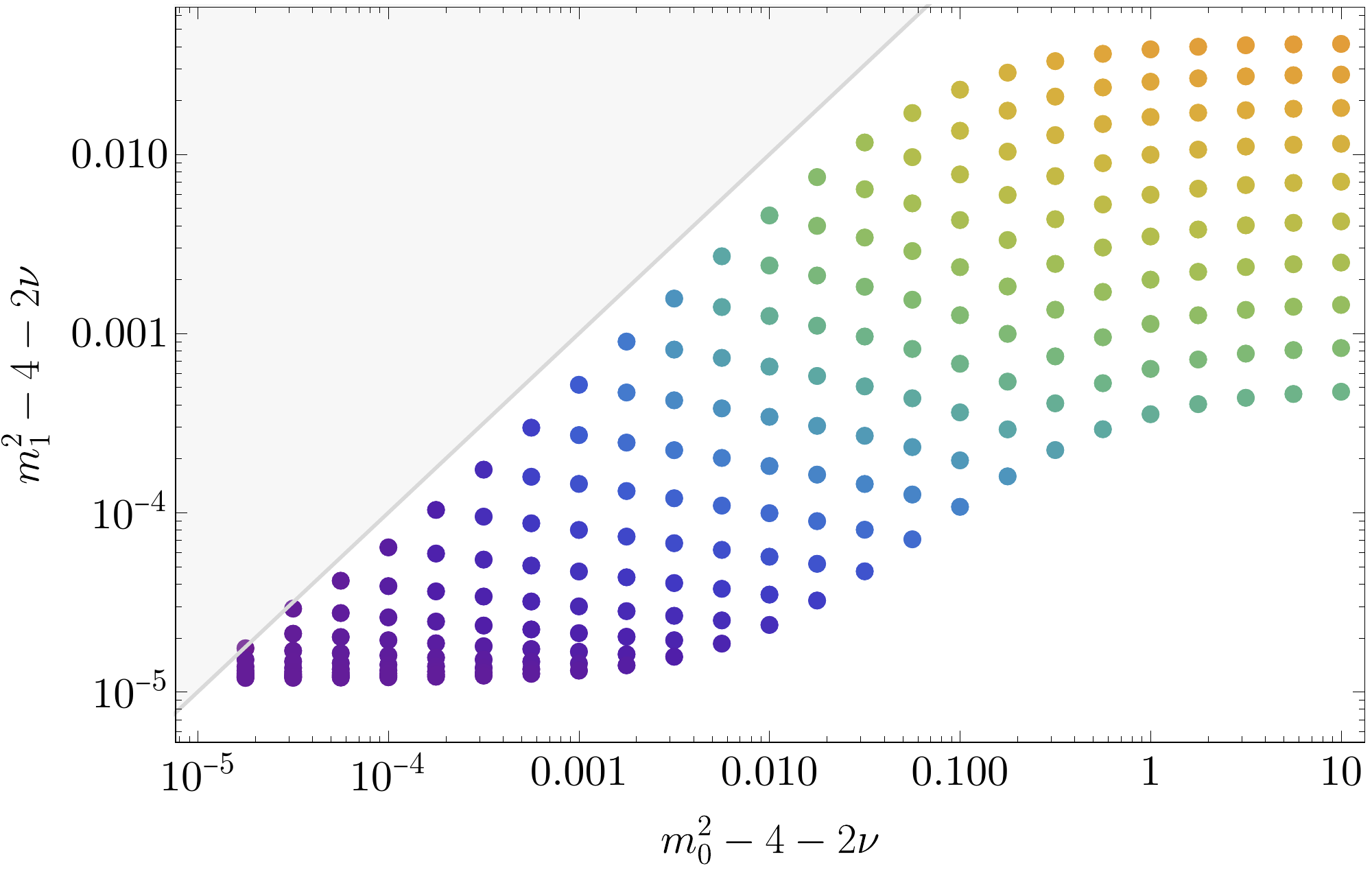}}
	\caption{These plots show the amount of tuning required between the Higgs and the GW sector when the breaking scale is set to $f=10\;\si{\tera\electronvolt}$ (left figure) and $f=50\;\si{\tera\electronvolt}$ (right figure). Different points which have the same $\alpha_0$ value are obtained by varying $\nu$ which is decreasing as one goes up on the vertical axis. We can observe that the tuning increases as we go the the top-right region of the $\alpha_0-\alpha_1$ plane. Also increasing the breaking scale increases the tuning, which is expected. }
	\label{fig:higgs-gw-tuning}
\end{figure}

Since the GW contribution dominates the effective potential, the minimum is mainly determined by the GW sector. Therefore, one can choose a particular value for the IR mistune $\delta T_1$, such that the minimum coincides with $y_1^{\text{c}}$, as long as \eqref{stability_condition} is satisfied. We denote this ``critical" mistune by $\delta T_1^{c}$. Then, a convenient parameter to measure the tuning between the GW and the Higgs sector is $\abs{\delta T_1/\delta T_1^c}-1$.

In Figure~\ref{fig:higgs-gw-tuning}, we show the required tuning on the parameter space where the electroweak symmetry breaking is activated by the radion. The procedure to obtain these plots is as follows: We have set the GW sector parameters to be $v_0 = 1/50$, $v_1=1$ and $\epsilon=1/10$. Then we set the conformal breaking scale to be $f=10(50)\;\si{\tera\electronvolt}$ for the left(right) plot. For each point on the $\alpha_0-\alpha_1$ plane, we fix the Higgs parameters such that $v_{\text{eff}}=246\;\si{\giga\electronvolt}$ and $m_h=126\;\si{\giga\electronvolt}$. Finally, we solve for $\delta T_1$ such that $V_{\text{eff}}^{\text{IR}}=0$.

Different points in Figure~\ref{fig:higgs-gw-tuning} which share the same $\alpha_0=m_0^2-4-2\nu$ values are obtained by varying $\nu$. For a fixed $\alpha_0$, larger $\nu$ values correspond to larger $\alpha_1=m_1^2-4-2\nu$, thus $\nu$ increases as one goes up on the vertical axis. As one can see more clearly in the plot on the right, tuning between the GW and the Higgs sectors does also increase in this direction. Therefore, as one relaxes the tuning in the $m_1^2$ parameter, the tuning in $\delta T_1$ becomes larger.

\subsubsection*{The Radion Mass}
\label{sec:radion_mass_gw}

To calculate the radion mass, we make the following ansatz for the radion wavefunction, and the radion mass eigenvalue:
\begin{align}
	\label{radion_ansatz}
	F_r=e^{2A}\left(1+\kappa^2 \tilde{F}_r\right)\qq{and}m_r^2=\kappa^2 l_r^2.
\end{align}
By plugging this ansatz into \eqref{eq:17}, and keeping only the terms which are at leading order in $\kappa^2$ we find
\begin{align}
\label{eq:radion_diff}
	\tilde{F}_r''+\tilde{F}_r'\left(2-2\frac{\phi''}{\phi'}\right)=\frac{2}{3} \left[(\phi'^2+v'^2)+\left(v''-\frac{\phi''}{\phi'}v'\right)h_r e^{-2y}\right]-e^{2y}l_r^2.
\end{align}
The solution of this equation is
\begin{align}
	\label{radion_sol}
	\tilde{F}_r'(y)=\frac{1}{u(y)}\left[\frac{2}{3}\int_0^y \dd{y'}u(y')(\phi'^2 + g_v(y'))-l_r^2 \int_0^y \dd{y'}u(y')e^{2y}+\tilde{F}_r'(0)\right],
\end{align}
where
\begin{align}
	u(y)=\exp{\int_0^y \dd{y'}\left(2-2\frac{\phi''}{\phi'}\right)}\qq{and}g_v(y)=v'^2+\left(v''-\frac{\phi''}{\phi'}v'\right)h_r e^{-2y}.
\end{align}
The function $u(y)$ can be obtained analytically:
\begin{align}
	u(y)=\left(\frac{\phi'(0)}{\phi'(y)}\right)^2 e^{2y}.
\end{align}
By plugging this result into \eqref{radion_sol} and evaluating it at $y=y_1$, we can write an expression for the radion mass:
\begin{align}
	\label{radion_mass_1}
	\frac{l_r^2}{f^2}=\left(\int_0^{y_1} \dd{y}\frac{e^{4(y-y_1)}}{\phi'(y)^2}\right)^{-1}\left[\frac{1}{3}+\frac{2}{3}\int_0^{y_1}\dd{y}\frac{g_v(y)}{\phi'^2}e^{2(y-y_1)}+\left(\frac{\tilde{F}'(0)e^{-2 y_1}}{\phi'(0)^2}-\frac{\tilde{F}'(y_1)}{\phi'(y_1)^2 }\right)\right],
\end{align}
where we have dropped the $e^{-2 y_1}$ term after evaluating the integral $\int_0^{y_1} \dd{y}e^{2(y-y_1)}$. The boundary values for $\tilde{F}_r'$ can be expressed in terms of the boundary values for $h_r$ using the junction conditions given in \eqref{eq:11}. Since $\varphi=0$ on the branes in the case of stiff wall boundary conditions, these give
\begin{align}
	\tilde{F}_r'=\frac{1}{3}v' h_r e^{-2y},\qq{on the branes.}
\end{align}
Plugging this result into \eqref{radion_mass_1} gives
\beq
\begin{aligned}
	\label{radion_mass_gw_h}
		\frac{l_r^2}{f^2}&=\left(\int_0^{y_1} \dd{y}\frac{e^{4(y-y_1)}}{\phi'^2}\right)^{-1}\biggl[\frac{1}{3}+\frac{2}{3}\int_0^{y_1}\dd{y}\frac{g_v(y)}{\phi'^2}e^{2(y-y_1)}\\
		&\phantom{{}={}}+\frac{1}{3}\left(\frac{v'(0)h_r(0)}{\phi'(0)^2}-\frac{v'(y_1)h_r(y_1)}{\phi'(y_1)^2 }\right)e^{-2 y_1}\biggr].
\end{aligned}
\eeq
The first term in this expression is the radion mass in the absence of the Higgs VEV, which can be calculated analytically. Ignoring the terms which are proportional to $e^{-2 y_1}$ and higher, it is given by
\begin{align}
	\label{radion_mass_gw}
	\frac{l_{r,0}^2}{f^2} \equiv \frac{1}{3}\left(\int_0^{y_1} \dd{y}\frac{e^{4(y-y_1)}}{\phi'^2}\right)^{-1}=\frac{4}{3}\epsilon (2-\epsilon)^2 v_0^2 e^{\epsilon y_1} \left[\frac{(v_1/v_0)(4-\epsilon)}{4-2\epsilon}-e^{\epsilon y_1}\right].
\end{align}
We note that this result confirms the stability condition we have derived in \eqref{stability_condition}.

\begin{figure}
	\centering
	\subfloat{\includegraphics[width=0.48\textwidth,keepaspectratio]{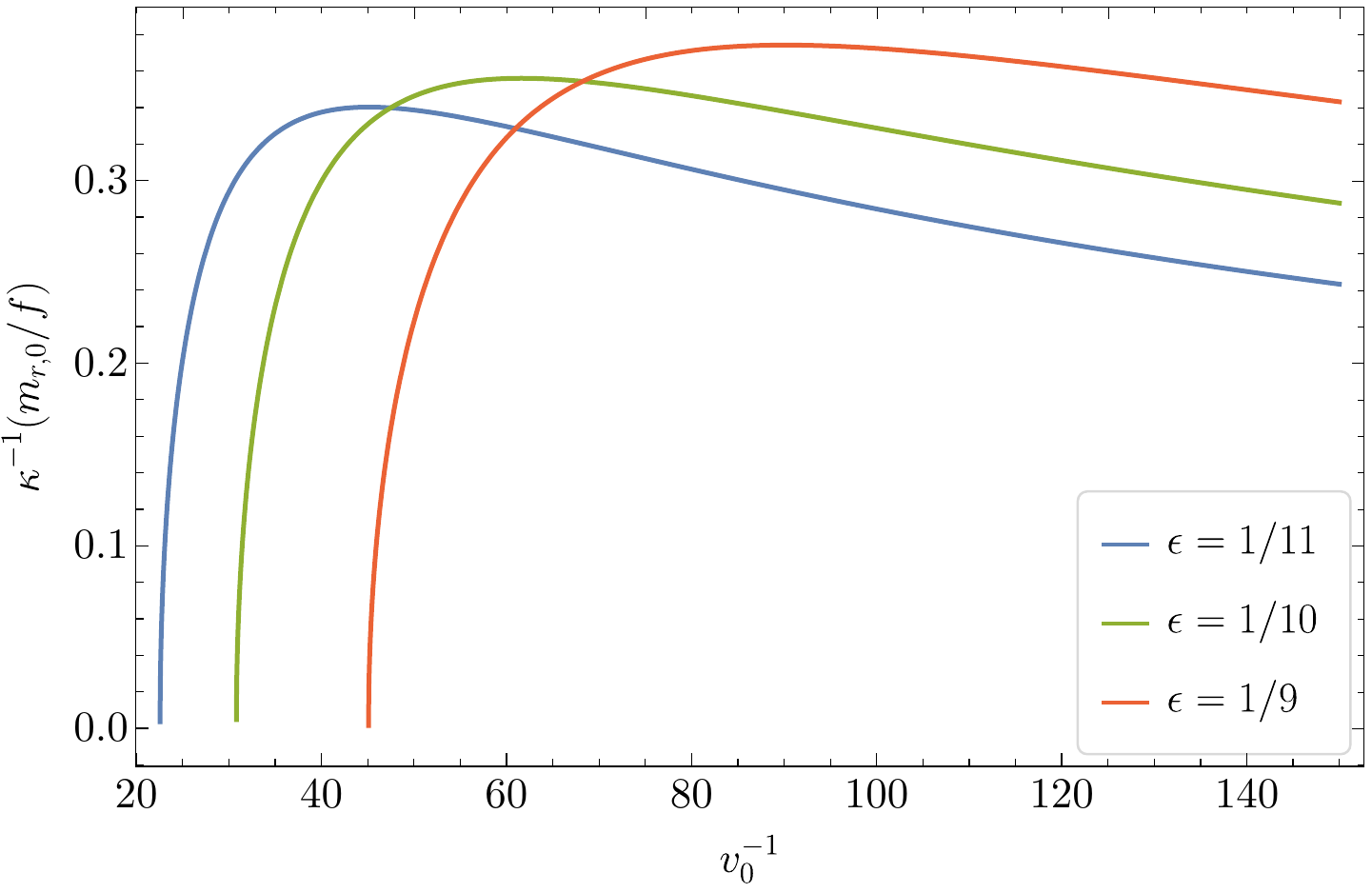}}\hfill%
	\subfloat{\includegraphics[width=0.48\textwidth,keepaspectratio]{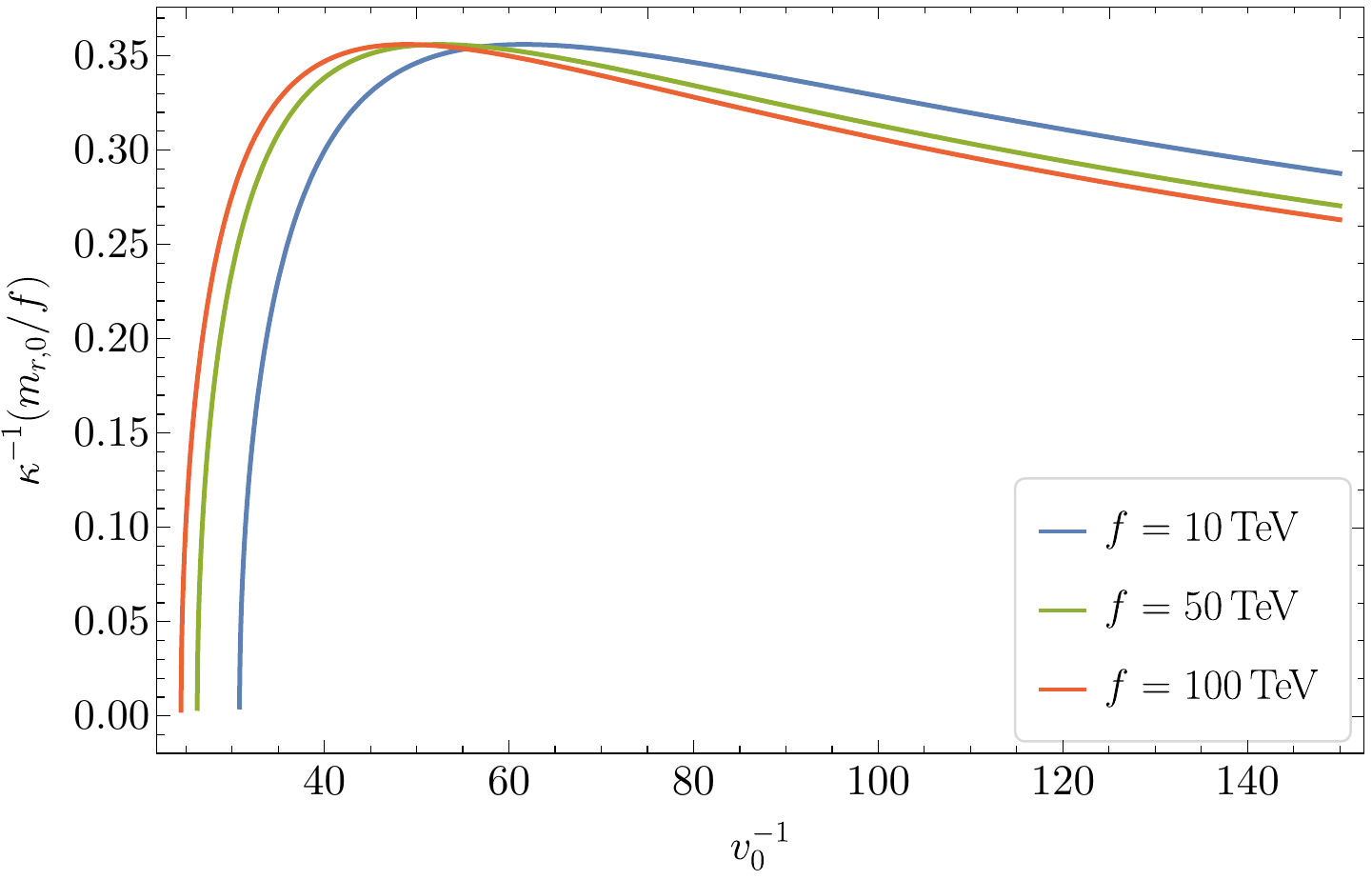}}
	\caption{In these plots, we show the values of the radion mass rescaled with the breaking scale $f$ and backreaction parameter $\kappa$. In both plots, we fix $v_1 = 1$ and vary $v_0$. On the left figure, we fix the breaking scale to $f=10\,\si{\tera\electronvolt}$ and plot for various values of $\epsilon$. On the right figure, we fix $\epsilon=1/10$ and plot for various values of the breaking scale. }
	\label{fig:radion-mass-gw}
\end{figure}

In the case of nonzero Higgs VEV, the radion mass receives contributions due to its mixing with the Higgs, which are given by the remaining terms in \eqref{radion_mass_gw_h}. To calculate them, all we need is the profile for the field $h_r$. It can be obtained by solving \eqref{eq:13} together with the boundary condition \eqref{eq:15} in the small backreaction limit. By using the ansatz \eqref{radion_ansatz}, the equation of motion and the boundary condition for $\tilde{h}_r \equiv h_r e^{-2 y}$ are given respectively by
\begin{align}
	\label{hr_eq}
	\tilde{h}_r''-\nu^2 \tilde{h}_r=4[3v'+(-4+\nu^2)v]=4(v''-v'),
\end{align}
and
\begin{align}
	\label{hr_bc}
	(\tilde{h}_r'+2 \tilde{h}_r)-2 v'=\pm \frac{1}{2}\pdv[2]{V_{0,1}}{v}\tilde{h}_r,\qq{$+/-$ on the UV/IR brane.}
\end{align}
This equation can be solved exactly, and its solution can be plugged into \eqref{radion_mass_gw_h} to determine the correction to the radion mass due to Higgs-Radion mixing. The expression for the correction is fairly complicated, but it can be easily seen that it scales with $v^2(y_1)\sim v_{\text{eff}}^2/f^2$, so they are heavily suppressed compared to the leading term inside the brackets of \eqref{radion_mass_gw_h}. Hence we conclude that, for the parameter space we are interested in, the radion mass is given by \eqref{radion_mass_gw} to a very good approximation. We are showing the radion mass as a function of GW parameters in Figure~\ref{fig:radion-mass-gw}.

\subsubsection*{The Validity of Small Backreaction}
\label{sec:small-backreaction}
At the beginning of this Subsection, we have ignored the backreaction of gravity when obtaining the solutions for the VEVs of both field, and we have used those solutions to calculate the mass spectrum. Now, we shall check the validity of this approximation. 

In this model, the field profiles in the bulk have different scales. While the GW field starts with $\mathcal{O}(0.01)$ and ends with $\mathcal{O}(1)$, the Higgs field scales as $\sim v(y_1)e^{-2 y_1}$ and $\sim v(y_1)$ near the UV and IR branes respectively. Therefore, the backreaction of the GW field on the Higgs profile can be sizable, due to the fact that the GW field heavily dominates throughout the bulk. Hence, we should analyze the backreaction carefully, and determine the largest value of $\kappa$, so that the results we have derived so far remain valid. 

In order to study the backreaction, we need to use the general form of the equation of motion which is given by \eqref{genbulkeom}. The function $A'(y)$ is given by \eqref{ay_profile}, which for this model reads
\begin{align}
	A'=\sqrt{1+\frac{\kappa^2}{12}\left[\phi'^2 + v'^2 - \epsilon(\epsilon -4)\phi^2-(-4+\nu^2)v^2\right]}.
\end{align}
We are only interested in the backreaction of the GW field to leading order. Therefore we approximate $A'$ as
\begin{align}
	A'\approx 1+\frac{\kappa^2}{24}\left(\phi'^2-\epsilon(\epsilon-4)\phi^2\right).
\end{align}
This function peaks on the IR brane, thus we can use its IR value to determine the strength of the backreaction. Hence we estimate the size of the corrections as
\begin{align}
	\frac{\kappa^2}{24}\left(\phi'^2(y_1)-\epsilon(\epsilon-4)\phi^2(y_1)\right)\approx \frac{\kappa^2}{24}\times \mathcal{O}(1),
\end{align}
Although this number seems quite small, it has a significant effect on the Higgs VEV as we shall see below.

The Higgs VEV on the IR brane is determined by the IR boundary condition for the Higgs, which can be rearranged to have the form
\begin{align}
	\label{higgs-ir-bc-alter}
	\frac{v^2(y_1)}{v_H^2}=1-\frac{2}{\lambda_H v_H^2}\frac{v'(y_1)}{v(y_1)}.
\end{align}
The Higgs parameters we have used in this Subsection typically have $\lambda_H^{-1}\approx 17$ and $m_1^2 \approx 4$ to satisfy $v_{\text{EW}}=246\,\si{\giga\electronvolt}$ and $m_{h}=126\,\si{\giga\electronvolt}$. This sets $v_H^2 = \lambda_H^{-1} m_1^2 \approx 68$. So the number on the LHS of \eqref{higgs-ir-bc-alter} is very small. Therefore, a delicate cancellation between two $\mathcal{O}(1)$ parameters is needed. This equation encodes the tuning required to get a Higgs VEV that is suppressed compared with the KK scale.  For a fixed set of 5D parameters, Equation~\eqref{higgs-ir-bc-alter} effectively gives the Higgs VEV as a function of $y_1$, which is the position of the IR brane that minimizes the effective potential.  One can alternatively see that there is a large sensitivity to 5D parameters.  For example, by increasing the value of the 5D Newton constant, $\kappa^2$ even slightly, one finds that the Higgs VEV will change significantly because of the sensitivity of the cancellation to small changes.

To see explicitly how large the backreaction parameter $\kappa$ can be, let us assume that we want to know $v^2(y_1)$ to a 10\% accuracy. Then a rough upper bound for $\kappa$ can be given by
\begin{align}
	\label{kappa-limit-gw}
	\frac{\kappa^2}{24}\left(\phi'^2(y_1)-\epsilon(\epsilon-4)\phi^2(y_1)\right)\lesssim 0.1\times \frac{v^2(y_1)}{v_H^2}.
\end{align}
For a hierarchy $f/v_{\text{eff}} \sim 10$, we get $\kappa \lesssim \mathcal{O}(1)\times 10^{-3}$.

In the next Subsection, we will study the same model but without the GW field. Then the only source of the backreaction is due to the Higgs field itself. In that case, we need to replace the GW field in the above expression by the Higgs field:
\begin{align}
	\label{kappa-limit-higgs}
	\frac{\kappa^2}{24}\left(v'^2(y_1)-(-4+\nu^2)v^2(y_1)\right)\lesssim 0.1\times \frac{v^2(y_1)}{v_H^2}.
\end{align}
The term inside the parentheses on the LHS is $\sim 8 v^2(y_1) $, thus the bound on $\kappa$ in this model is given by
\begin{align}
	\kappa \lesssim \sqrt{0.1\times \frac{3}{v_H^2}} \sim \mathcal{O}(1)\times 10^{-2}.
\end{align}
\begin{figure}
	\centering
	\subfloat{\includegraphics[width=0.48\textwidth,keepaspectratio,height=5cm]{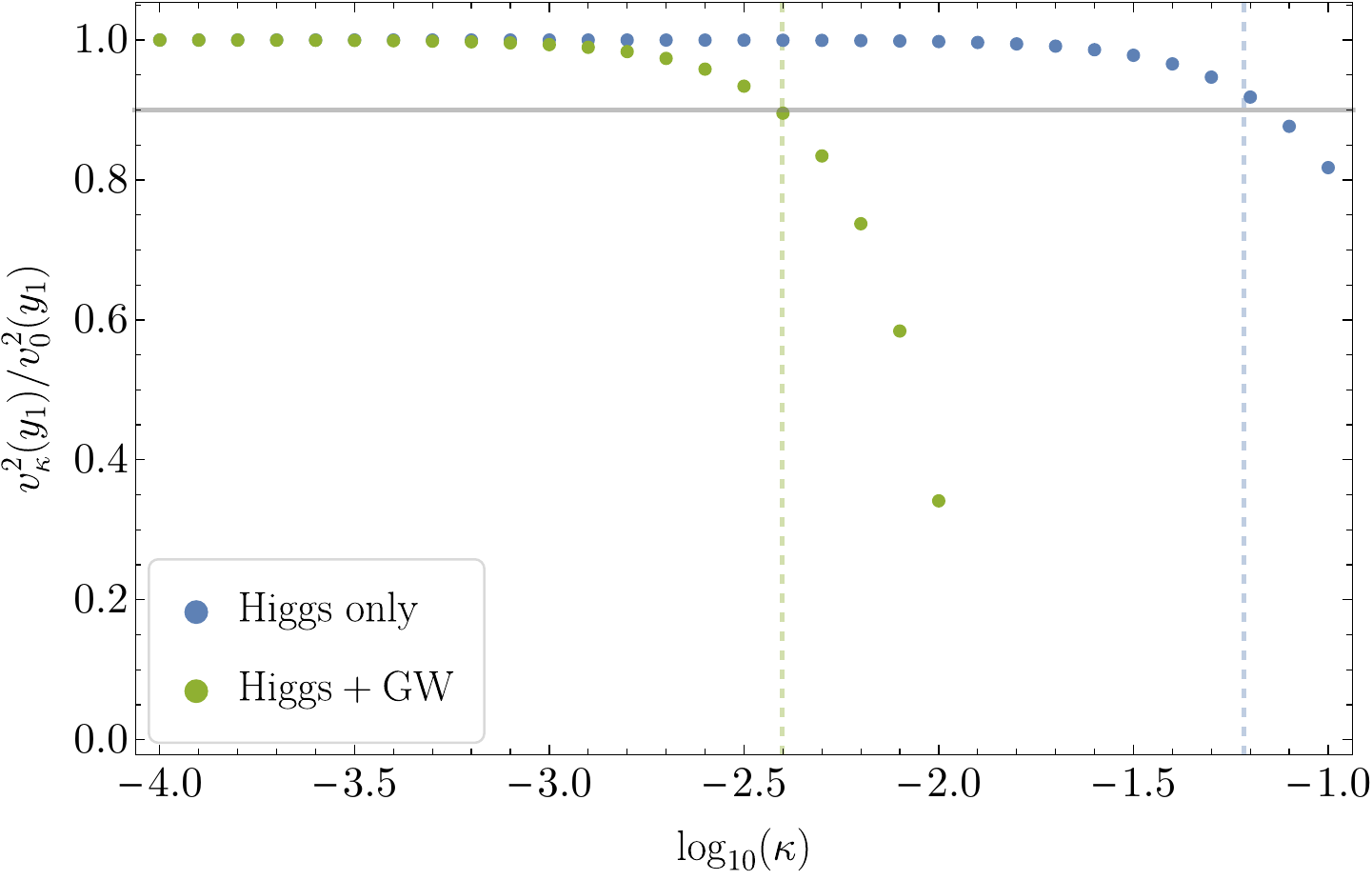}}\hfill%
	\subfloat{\includegraphics[width=0.48\textwidth,keepaspectratio,height=5cm]{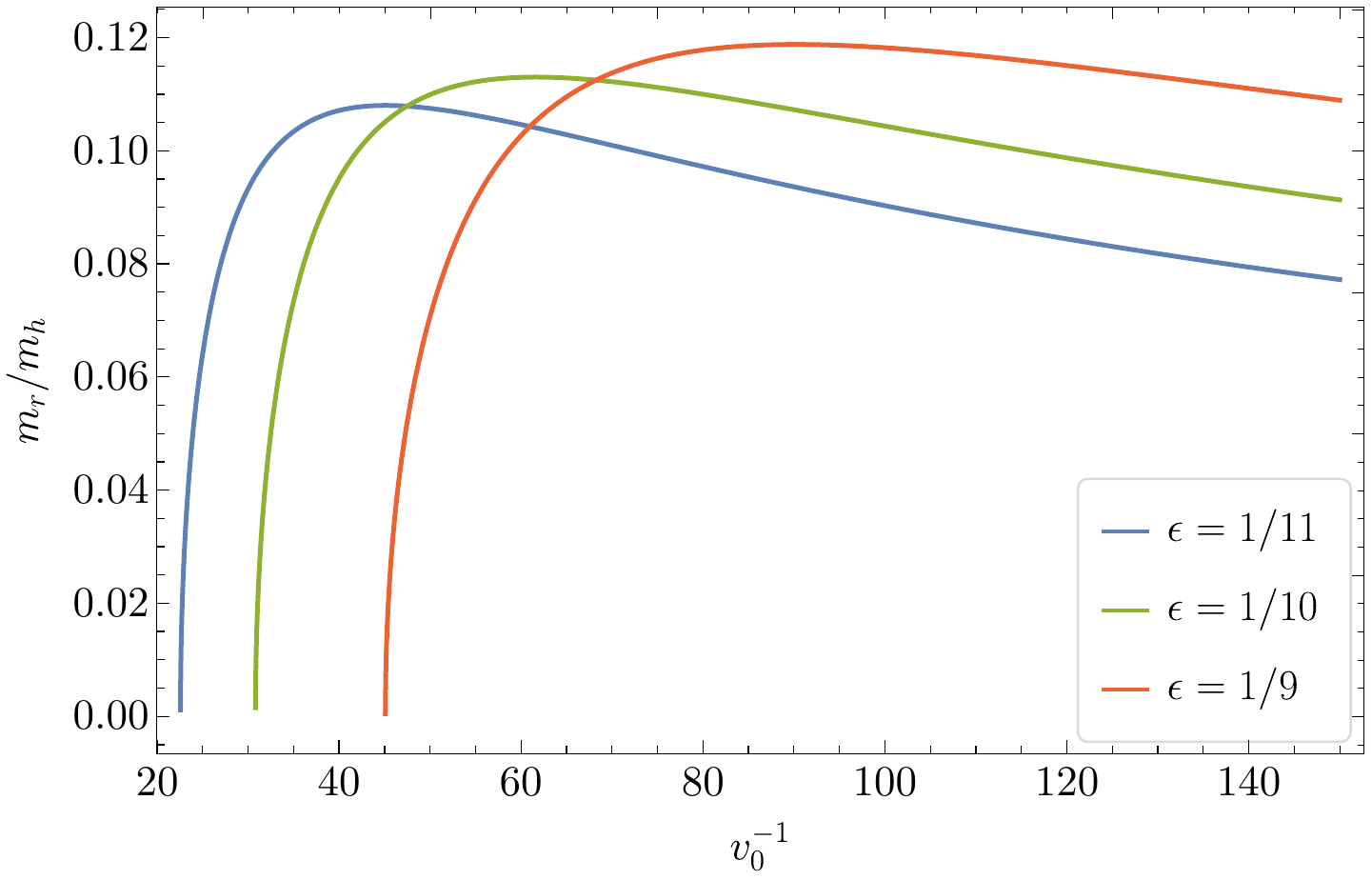}}
	\caption{On the left figure, we show the ratio between $v^2(y_1)$ obtained by solving \eqref{genbulkeom} numerically with nonzero backreaction, and the analytical result \eqref{higgs_vev_on_ir} obtained by neglecting the backreaction. The green dots show the results for the model of this Subsection, where both GW and the Higgs are in the bulk, while the blue dots are the results for the model of the next Subsection, where only the Higgs is in the bulk. The green and blue dashed vertical lines show the upper bound estimates for $\kappa$ calculated via \eqref{kappa-limit-gw} and \eqref{kappa-limit-higgs} respectively. The gray line corresponds to a ratio of $0.9$ which we have assumed in our estimates. On the right figure, we plot the Radion-Higgs mass ratio for $\kappa = 4\times 10^{-3}$. The parameters for these plots are $v_0=1/50$, $v_1=1$, $\epsilon=1/10$, $f=10\,\si{\tera\electronvolt}$, $m_0^2=43/10$, $\nu=1/10$ unless otherwise specified. The remaining free parameters are fixed such that $v_{\text{EW}}=246\,\si{\giga\electronvolt}$ and $m_h=126\,\si{\giga\electronvolt}$.}
	\label{fig:backreaction}
\end{figure}

On the left plot of Figure~\ref{fig:backreaction}, we show the ratio $v^2_{\kappa}(y_1)/v^2_{0}(y_1)$ as a function of $\kappa$, where $v^2_{\kappa}(y_1)$ is the Higgs VEV square on the IR brane found by numerically solving \eqref{genbulkeom} with nonzero backreaction, and $v^2_{0}(y_1)$ is the analytical result with no backreaction given in \eqref{higgs_vev_on_ir}. The green and blue dots show the results for models of this and the next Subsection respectively. The green and blue dashed vertical lines denote the bounds calculated via \eqref{kappa-limit-gw} and \eqref{kappa-limit-higgs} respectively, while the gray line shows the $v^2_{\kappa}(y_1)/v^2_{0}(y_1)$ ratio of $0.9$, which we have used in our estimates. We can see that the backreaction has a much bigger effect when both GW and the Higgs are in the bulk, and our estimates agree with the numerical results quite well. 

On the right plot of Figure~\ref{fig:backreaction}, we show the Radion-Higgs mass ratio by assuming $\kappa= 4\times 10^{-3}$. We see that, if we insist on staying in the regime where the backreaction can be neglected, then the radion mass is about an order of magnitude smaller than the Higgs mass. The radion can be made heavier by increasing the backreaction, but one needs a full numerical analysis of the mass spectrum, which is beyond the scope of this work. 

\subsection{Higgs in the Bulk}
\label{sec:higgs_only_in_the_bulk}

In this Subsection, we will study a model where the Higgs is the only stabilizing field in the bulk. This model has two major differences compared to the one in the previous Subsection where both the Goldberger-Wise and the Higgs fields were propagating in the bulk. First, the parameter space for electroweak symmetry breaking receives another constraint due to the fact that now the Higgs is responsible for radius stabilization. This means that we don't have the freedom of adjusting Goldberger-Wise parameters to get a stable minimum; instead, we demand that the Higgs parameters $\alpha_{0,1}$ and $\nu$ satisfy additional conditions which we shall derive below. The second main difference will be the radion mass, which now depends only on the Higgs parameters. 

\subsubsection*{Parameter Space for Electroweak Symmetry Breaking}

We will be using the notation introduced in Subsection~\ref{sec:ew_parameter_space_gw_higgs} and assume $\alpha_0>0$ so that \eqref{vev_derivative} is positive definite. From \eqref{higgs_eff_pot_final}, we can write the effective potential everywhere as
\begin{align}
	V_{\text{eff}}(y_1)=\begin{cases}
	e^{-4 y_1}\delta T_1 , &\text{No EWSB}\\
	e^{-4 y_1}\left[\delta T_1 - \frac{\lambda_H}{4}v^4(y_1)\right], &\text{EWSB},
	\end{cases}
\end{align}
where we have tuned $\delta T_0 = 0$. By taking the derivative and setting it to zero, we find that at an extremum point $y_1=y_1^{\text{ext}}$, the IR brane mistune is given by
\begin{align}
	\delta T_1^{\text{ext}}=\frac{\lambda_H}{8}v^2(y_1^{\text{ext}})\left[ 2 v^2(y_1^{\text{ext}})-\pdv{v^2(y_1)}{y_1}\eval_{y_1=y_1^{\text{ext}}}\right].
\end{align}
In order to not have a runaway solution in the UV, i.e.\ to have a global minimum, we demand that this term is positive. Then the second derivative of $V_\text{eff}$ at $y_1=y_1^{\text{ext}}$, where $\delta T_1$ is replaced by $\delta T_1^{\text{ext}}$ is given by
\begin{align}
	V_\text{eff}''(y_1^{\text{ext}})=e^{-4 y_1}\frac{\lambda_H}{2}\left[v^2(y_1)\left(4 \pdv{v^2(y_1)}{y_1}-\pdv[2]{v^2(y_1)}{y_1}\right)-\left(\pdv{v^2(y_1)}{y_1}\right)^2\right]\eval_{y_1=y_1^{\text{ext}}}.
\end{align}
By using explicit expression for $v^2(y_1)$, it is possible to show that $\delta T_1^{\text{ext}}>0$ implies $V_\text{eff}''(y_1^{\text{ext}})>0$, so we can use the former. Then we obtain the following constraint on the parameter $\alpha_1$:
\begin{align}
	\label{alpha-1-constraint}
	\alpha_1 > \frac{4\alpha_0 \nu \left[(1+\nu)(\alpha_0+4\nu)e^{2\nu y_1}-\alpha_0\right]}{[(e^{2\nu y_1}-1)\alpha_0 + 4\nu e^{2\nu y_1}]^2}\eval_{y_1=y_1^{\text{ext}}}.
\end{align}
 
\begin{figure}
	\centering
	\subfloat{\includegraphics[width=0.48\textwidth,height=5cm]{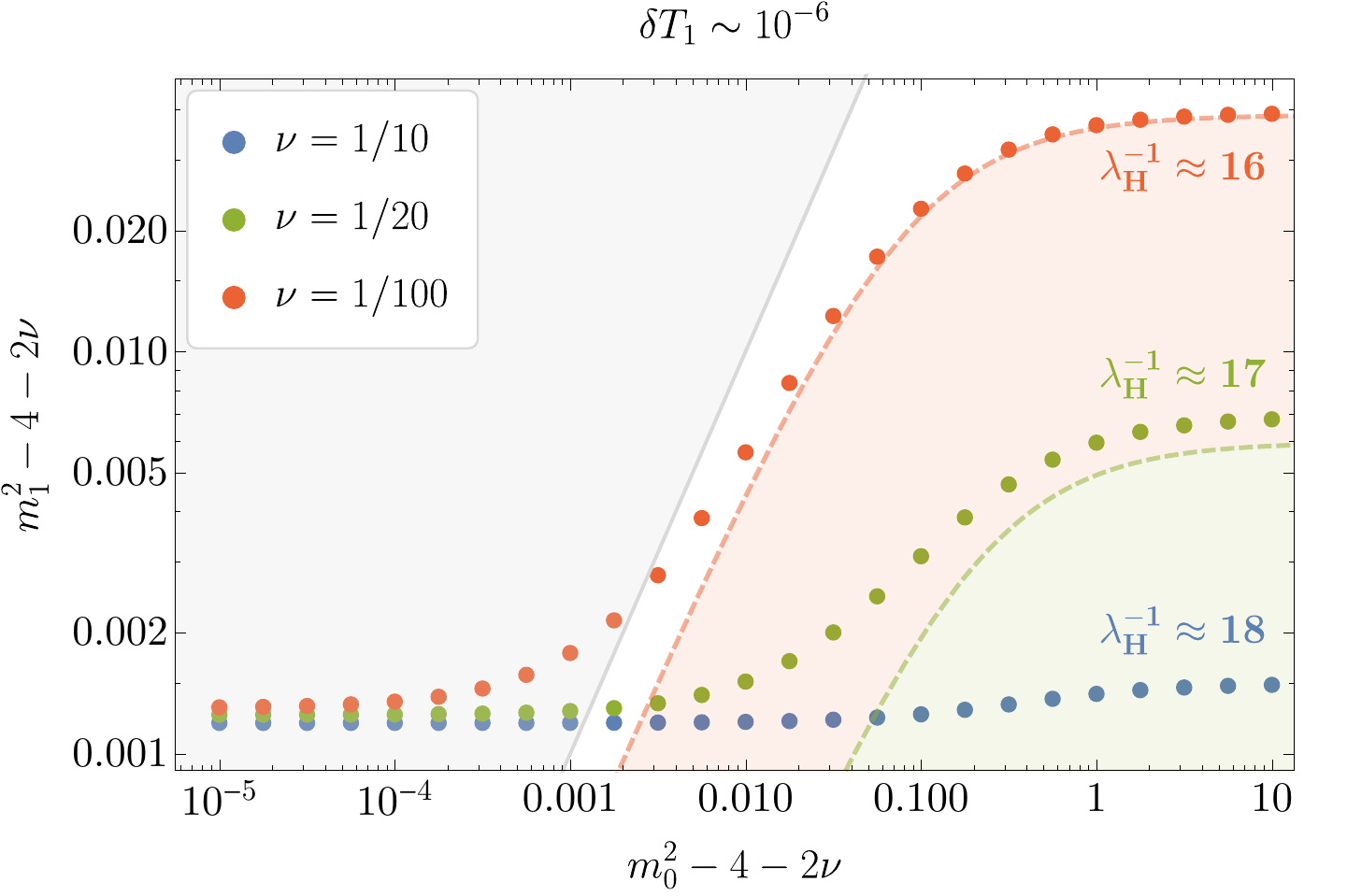}}\hfill%
	\subfloat{\includegraphics[width=0.48\textwidth,height=5cm]{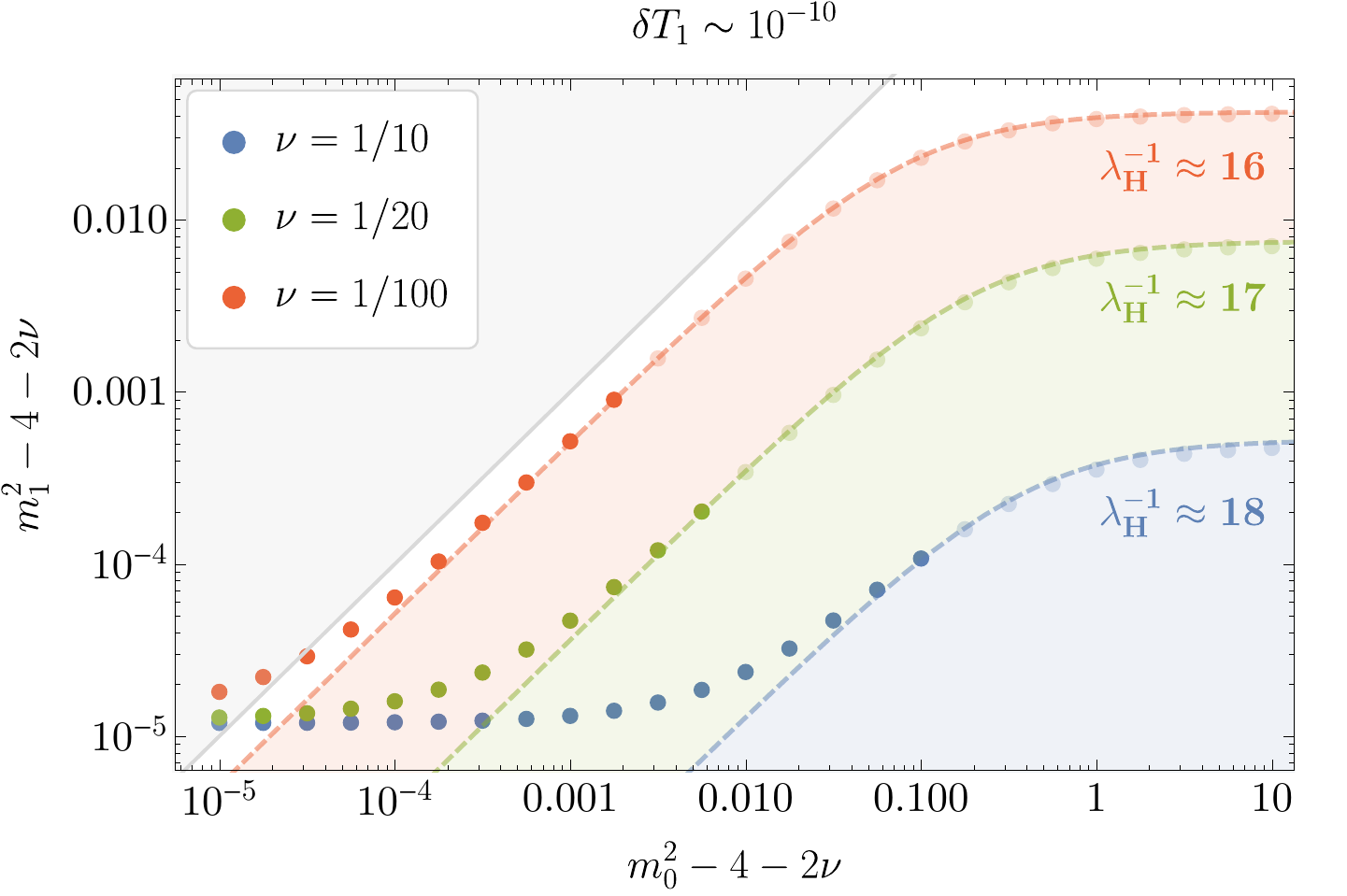}}
	\caption{In this figure, the circles denotes possible $\qty{\alpha_0, \alpha_1}$ pairs characterized by $v_{\text{EW}} = 246\,\si{\giga\electronvolt}$ and $m_h=126\,\si{\giga\electronvolt}$. The hierarchy between the electroweak and the conformal breaking scales is $f/v_{\text{EW}}=10$ and $f/v_{\text{EW}}=100$ for the left and right plot respectively. The gray shaded area denotes the region in parameter space where there is no ``critical" value of $y_1$. The color shaded areas are the regions where the condition for global minimum \eqref{alpha-1-constraint} is violated. To improve clarity, we are showing the $\qty{\alpha_0, \alpha_1}$ pairs corresponding to these regions with semi-transparent circles. }
	\label{fig:higgs-tuning}
\end{figure}

We are showing the effect of this constraint on the Higgs parameter space in Figure~\ref{fig:higgs-tuning}. In these plots, we are plotting $m_1^2$ in terms of $m_0^2$ such that the electroweak scale is $v_{\text{EW}} = 246\,\si{\giga\electronvolt}$ and the Higgs mass is $m_h=126\,\si{\giga\electronvolt}$. The methods and the equations for determining $m_1^2$ are identical to the Goldberger-Wise and Higgs model studied in the previous Subsection. The semi-transparent circles correspond to the points where the constraint \eqref{alpha-1-constraint} is not satisfied, hence those points should be excluded from the parameter space. We can see that the constraint becomes more stringent as one decreases $\nu$ and increases the breaking scale $f$. 

\subsubsection*{The Radion Mass}

\begin{figure}
	\centering
	\subfloat{\includegraphics[width=0.48\textwidth,height=5cm]{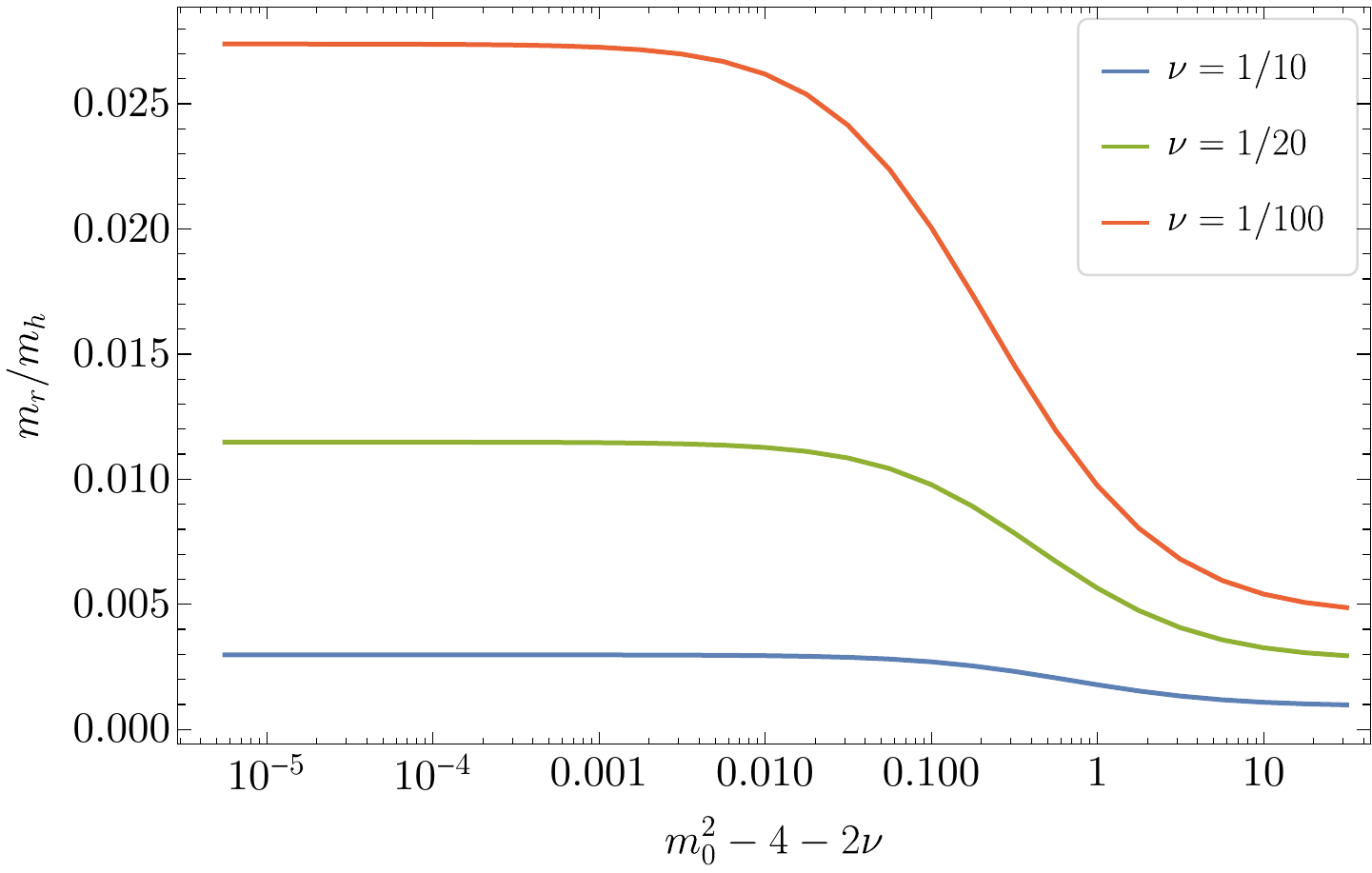}}\hfill%
	\subfloat{\includegraphics[width=0.48\textwidth,height=5cm]{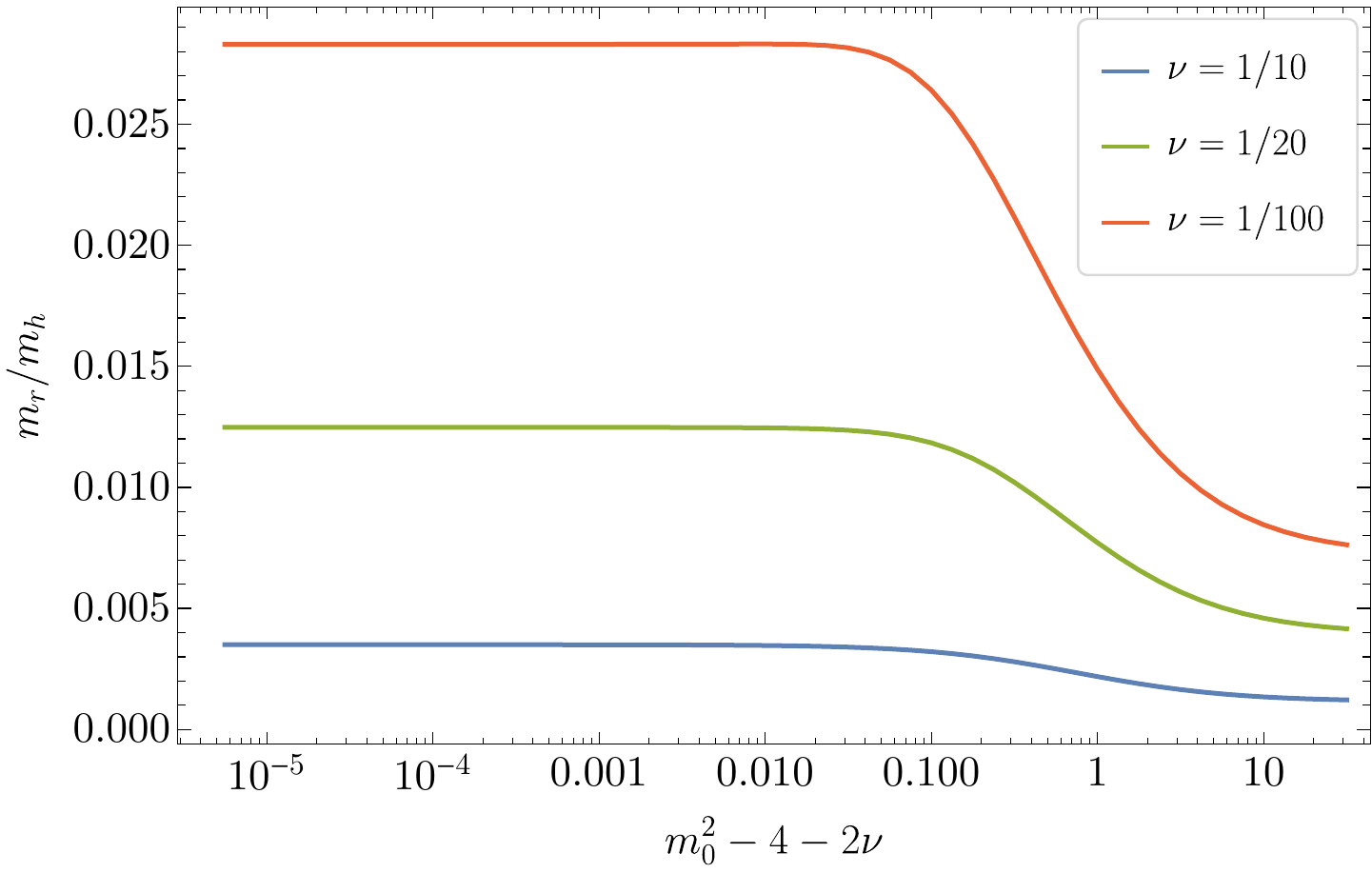}}
	\caption{In this figure, we are plotting the ratio between the radion mass and the Higgs mass as a function of $\alpha_0$ by assuming $\kappa=6\times 10^{-2}$. The breaking scale is $f=10\,\si{\tera\electronvolt}$ and $f=50\,\si{\tera\electronvolt}$ on the left and right figure respectively. The Higgs parameters are fixed such that $v_{\text{EW}} = 246\,\si{\giga\electronvolt}$ and $m_h = 126\,\si{\giga\electronvolt}$. }
	\label{fig:radion-higgs-mass}
\end{figure}

The radion mass can be calculated by following the procedure described in Subsection~\ref{sec:radion_mass_gw}. For this model we find
\begin{align}
	\label{radion_higgs_mass_exp}
	\frac{l_r^2}{f^2}=\frac{1}{3}\left(\int_0^{y_1} \dd{y}\frac{e^{4(y-y_1)}}{v'(y)^2}\right)^{-1}\left[1-\frac{\tilde{h}_r(y_1)}{v'(y_1)}+\frac{\tilde{h}_r(0)}{v'(0)}e^{-2 y_1}\right],
\end{align}
where $\tilde{h}_r$ is the solution to equations \eqref{hr_eq} and \eqref{hr_bc}. Again, the full solution is analytical, but fairly complicated. Nevertheless, we can get a good insight by calculating the factor which multiplies the term inside the brackets. It is given by
\begin{align}
	I\equiv \frac{1}{3}\left(\int_0^{y_1} \dd{y}\frac{e^{4(y-y_1)}}{v'(y)^2}\right)^{-1}=\frac{4 v^2(y_1) m_0^2 \nu^2 e^{-2\nu y_1}}{3(e^{2\nu y_1}-1)(m_0^2-4+2\nu)}\left[\frac{e^{2\nu y_1}(2+\nu)-r(2-\nu)}{(1-r e^{-2\nu y_1})^2}\right],
\end{align}
where $r = \frac{m_0^2-4-2\nu}{m_0^2-4+2\nu}$. We find that the term inside the brackets varies between $1.0$ and $1.4$ in the parameter space we are using. We show the results of the Radion-Higgs mass ratio in Figure~\ref{fig:radion-higgs-mass} as a function of $\alpha_0 = m_0^2-4-2\nu$, under the assumption that $\kappa = 6\times 10^{-2}$. To make the plots we have used the full analytical result given in \eqref{radion_higgs_mass_exp}. We can see that in this model, the radion is two to three orders of magnitude lighter than the Higgs. Since the effect of the backreaction is not as significant as in the previous model, we expect that a significantly light radion is a property of this model, even with backreaction.\footnote{Note that radion masses at the GeV scale and above are still largely unconstrained by the latest experimental results, as shown by the detailed analysis of~\cite{Abu-Ajamieh:2017khi}.} 

It might seem surprising that decreasing $\nu$ increases the radion mass. To understand the effect of this parameter, it is useful to study the quantity $I$ in detail. Since $f\sim \mathcal{O}(10\,\si{\si\tera\electronvolt})$, we have $y_1 \sim \mathcal{O}(30)$. Choosing $\nu \approx 1/10$ implies $e^{2\nu y_1}\sim \mathcal{O}(1000)$ hence we can assume $e^{2\nu y_1}\gg 1$. In this approximation, $I$ becomes
\begin{align}
	I\approx \frac{4}{3}v^2(y_1) \frac{m_0^2 (2+\nu)\nu^2}{m_0^2-4+2\nu}e^{-2\nu y_1},\quad e^{2\nu y_1}\gg 1.
\end{align}
In this regime, the radion mass is heavily suppressed by $e^{-2\nu y_1}$, which is $\mathcal{O}(10^{-3})$, in addition to suppression by $v^2(y_1)$ and $\nu^2$. On the other hand, if $\nu \sim 1/100$, then $e^{2\nu y_1}\sim \mathcal{O}(1)$. In this case, we can evaluate $I$ at the leading order in $\nu$. The result is
\begin{align}
	I \approx \frac{8}{3}v^2(y_1) \frac{m_0^2 \nu^2}{m_0^2-4}\frac{e^{2\nu y_1}}{(e^{2\nu y_1}-1)^2},\quad \nu \ll 1.
\end{align}
We see that the exponential suppression is absent in this case, hence the radion is heavier in this regime.


\section{CFT Interpretation}
\label{sec:cftinterpretation}
Field theories in AdS admit an interpretation in terms of a 4D strongly coupled CFT (or approximate CFT) dual~\cite{Maldacena:1997re,Witten:1998qj}.  In this Section, we discuss a CFT interpretation of multi-scalar stabilization models.

It is useful to first recall the CFT interpretation of the original unstabilized Randall-Sundrum construction~\cite{Randall:1999ee}, and then the simplest stabilization mechanism introduced by Goldberger and Wise~\cite{Goldberger:1999uk,Goldberger:1999un}.  We then discuss the specific interpretation of multi-scalar models with emphasis on those where there are critical points for symmetry breaking in the moduli space for the radion.

The original 2-brane RS model with weak 5D gravity has an interpretation as a large-$N$ approximate CFT where the conformal symmetry is spontaneously broken.  The radion degree of freedom corresponds to the dilaton -- the Goldstone boson of spontaneously broken conformal invariance.  The classical equations of motion (Einstein's equations) rule out a static geometry unless the UV and IR brane tensions are individually tuned against the bulk cosmological constant.  These two tunings represent (respectively) the need to have a vanishing bare cosmological constant so that there is no explicit breaking of conformal invariance, and the vanishing of the dilaton quartic, an allowed coupling that would otherwise destabilize the conformally non-invariant vacuum state.  In other words, as is well known~\cite{Fubini:1976jm}, spontaneous breaking of exact conformal invariance without breaking Poincar\'e invariance requires a flat direction.

Even in this case, with both tunings performed (and ignoring quantum effects)\footnote{In fact, even with both UV and IR brane tunings performed, quantum corrections will generally lead to a nonzero Casimir potential between the branes.}, the continuum of physically inequivalent vacua (the moduli space of inter-brane separations) present a problem for creating a low energy description that resembles our Standard Model.  Thus, a model must incorporate a stabilization mechanism.  The simplest such mechanism posits a bulk scalar field that develops a vacuum expectation value, deforming the geometry through gravitational backreaction.  If this scalar has an approximate shift symmetry, violated by (for example) a small bulk mass term, its VEV grows slowly, and backreaction effects are sizable only deep into the bulk of AdS.  This backreaction leads to an effective potential for the inter-brane separation, and stabilization is achieved dynamically without tuning of the IR brane tension (the UV tuning must still be performed, or enforced by additional symmetries to obtain static geometries).  The slow growth of the scalar VEV in the UV region of the geometry ensures that the generated hierarchy is exponentially large.  

This 5D picture is dual to 4D dimensional transmutation in which a nearly marginal operator (not necessarily classically marginal as is the case for asymptotic freedom) is given a nonzero coupling constant.  Its logarithmic evolution breaks conformal invariance explicitly, leading to the dynamical generation of an infrared scale where either the theory has flowed very far away from the UV fixed point~\cite{ArkaniHamed:2000ds,Rattazzi:2000hs}, or has flowed to a point where the effective dilaton quartic has become small~\cite{Bellazzini:2013fga,Coradeschi:2013gda}.

There is a similar class of 5D models where, in the 4D CFT dual, it is a composite operator that serves as the nearly marginal deformation of the CFT.   This occurs in the case where there is a 5D field with mass near the Breitenlohner-Freedman bound, which (in 5D) is $m^2 = -4$ in units of the AdS curvature~\cite{Klebanov:1999tb,Kaplan:2009kr,Gorbenko:2018ncu}.  The corresponding operator ${\mathcal O}$ in the CFT dual has dimension $\Delta \sim 2$, such that at large $N$, the composite operator ${\mathcal O}^\dagger {\mathcal O}$ has dimension $\Delta \sim 4$.  Thus, turning on this composite operator corresponds at large $N$ to a nearly marginal deformation of the CFT.  The model we discuss where only the Higgs is in the bulk is of this form, where it is the Higgs vacuum expectation value that backreacts on the geometry and stabilizes the inter-brane separation.

To understand the CFT interpretation of radion-induced Higgs criticality, it is first necessary to review the interpretation of the UV brane and field dynamics there.  In models with a UV brane, the interpretation is that there are two sectors of the theory, a fundamental sector and a CFT sector, and that these sectors mix.  

First, we review the basics of the correspondence without the UV brane (or with the UV brane taken to the boundary of AdS).  In this case the correspondence links field configurations on the boundary of AdS (which are not integrated over in the higher dimensional partition function) to sources for a dual CFT:
\begin{equation}
Z_\text{CFT} [ J(x) = \phi (x, z=0) ] = \int [{\mathcal D} \phi]_\text{bulk} \exp \left[ i S(\phi) \right].
\end{equation}
The CFT with sources could be described by a Lagrangian of the form ${\mathcal L}_\text{CFT} = {\mathcal L}^0_\text{CFT} + J (x) {\mathcal O}_\text{CFT}$.

When the UV brane is introduced, the path integral has no restriction -- effectively the sources are promoted to dynamical fields that mix with the CFT, and the new dual 4D Lagrangian is of the form
\begin{equation}
{\mathcal L} = {\mathcal L}_\text{fundamental}(\phi(x)) + {\mathcal L}^0_\text{CFT} + \lambda \phi(x) {\mathcal O}.
\end{equation}
The CFT and fundamental sector mix with each other.  Through the mixing,  spontaneous breaking of conformal invariance in the CFT sector from VEVs of CFT operators, $\langle {\mathcal O} \rangle$, is communicated to the fundamental sector.  The size of these VEVs corresponds in 5D to the position of the IR brane.  Integrating out the CFT (or bulk, in the 5D picture) degrees of freedom leads to an effective theory for the fundamental degrees of freedom in which parameters of the EFT, like the Higgs mass squared term, are a function of $y_1$.  This $y_1$ dependence can be such that the EFT crosses phase boundaries for symmetry breaking transitions as $y_1$ is varied.

In short, the 5D theory with two branes is dual to a 4D theory where a fundamental and CFT sector are coupled to each other, and where the CFT is spontaneously broken by operator VEVs.  The coupling between the two allows a ``CFT-induced criticality", where CFT operator VEVs induce instabilities in the fundamental sector.  In 5D, the instabilities leading to non-trivial VEVs are induced by varying the position of the IR brane.

In the case of using this picture as a model for electroweak physics, the effective Higgs mass squared term is, in the 4D picture, a function of CFT operator VEVs.  For certain values of these VEVs, the effective mass squared may become negative, and induce electroweak breaking.  In 5D, the mass squared of the lowest lying state of the bulk Higgs can become negative for ranges of $y_1$, such that electroweak symmetry breaking is induced by the radion VEV.  


\section{Conclusions}
\label{sec:conclusions}
We have explored radius stabilization in the context of 5D models in warped space with multiple scalar fields.  General results for multi-scalar models are derived to all orders in backreaction for 5D Einstein-Hilbert gravity.  We suggest a holography-inspired approach to the multi-scalar potential, and additionally derive a superpotential method for creating static geometries with multiple bulk scalar fields.

Of particular interest given recent experimental results are the phenomenological implications of Higgs fine tuning.  We explored 5D models focusing on obtaining a hierarchy between extra-dimensional excitations and the electroweak symmetry breaking scale.  5D models with such a hierarchy are close to the critical point for the electroweak sector, and thus, generically, the radion will be a dynamical degree of freedom that scans the Higgs phase transition.

Three examples were presented where the radion scans the effective Higgs mass parameter across a symmetry breaking phase transition:
\begin{itemize}
\item  A model with the Higgs on the IR brane, where a bulk Goldberger-Wise scalar couples to the Higgs and scans its effective mass.  
\item A model with only a Higgs field in the bulk, where coupling to the radion arises through geometrical backreaction.
\item A model with both Higgs and a Goldberger-Wise scalar in the bulk.
\end{itemize}
These three models cover a range of types of modulus-Higgs potential that give a hierarchy between extra-dimensional resonances and the electroweak scale.  

The models presented are an important step in understanding electroweak symmetry breaking in the context of holographic composite Higgs models, particularly those where the Higgs is light in comparison to the compositeness scale.  An important message is that the tuning of the Higgs itself has implications in terms of cosmology and collider phenomenology through the manner of the Randall-Sundrum/electroweak early universe physics and the low lying spectrum of scalar states.  

The formalism we have developed for multi-scalar stabilization models is relevant for all theories in which the Higgs emerges as a mode that is light in comparison with the compactification scale.  In general, the Higgs plays a non-trivial part in stabilization, and this work helps further elucidate the connection between the compactification scale and that of electroweak physics.


\section{Acknowledgements}
The authors thank Brando Bellazzini, Csaba Cs\'aki, Michael Geller, and Anson Hook for useful conversations regarding the research presented here.
The authors thank Cornell University for hospitality throughout the duration of this project.  JH thanks the Munich Institute for Astro- and Particle Physics (MIAPP) of the DFG cluster of excellence ``Origin and Structure of the Universe" for a productive research environment during a portion of this work.
CE, JH, and GR were supported in part by the DOE under grant award number DE-SC0009998.  CE is supported by the Deutsche Forschungsgemeinschaft under Germany's Excellence Strategy -- EXC 2121 ``Quantum Universe" -- 390833306.

\appendix
\section{Derivative of the Radion Potential}
\label{sec:derivativederivation}
In this Appendix, we derive Equation~\eqref{eq:dVbydy1} for the derivative of the effective potential.

Before imposing any of the boundary conditions, we straightforwardly take the derivative of the effective potential with respect to $y_1$:
\beq
\begin{aligned}
\dv{V_\textnormal{eff}}{y_1} &= \sum_i e^{-4 A} \left[ \frac{\partial V_0}{\partial \phi_i} \dv{ \phi_i}{ y_1} -\frac{1}{2 A'} \left( \phi'_i \dv{\phi'_i}{ y_1} - \frac{\partial V}{\partial \phi_i} \dv{ \phi_i}{ y_1} \right) \right] \eval_{y=y_0}  \\
&\phantom{{}={}}+  \sum_i e^{-4 A} \left[ \frac{\partial V_1}{\partial \phi_i} \dv{ \phi_i}{ y_1} +\frac{1}{2 A'} \left( \phi'_i \dv{\phi'_i}{ y_1} - \frac{\partial V}{\partial \phi_i} \dv{\phi_i}{y_1} \right) \right] \eval_{y=y_1}  \\
&\phantom{{}={}}-  4 e^{-4 A_1} \dv{A_1}{y_1} \tilde{V}_\text{IR} -  4 e^{-4 A_0} \dv{A_0}{y_1} \tilde{V}_\text{UV}.
\end{aligned}
\eeq
Imposing the bulk scalar equations of motion allows a substitution for $\partial V/\partial \phi_i$:
\beq
 \frac{\partial V}{\partial \phi_i} = \phi_i''-4 A' \phi_i'.
\eeq
Plugging this into the above derivative, we find:
\beq
\begin{aligned}
\dv{V_\textnormal{eff}}{y_1} &= \sum_i e^{-4 A} \left[ \left( \frac{\partial V_0}{\partial \phi_i}-2 \phi'_i \right) \dv{\phi_i}{y_1} -\frac{1}{2 A'} \left( \phi'_i \dv{\phi'_i}{ y_1} -\phi''_i \dv{\phi_i}{y_1} \right) \right] \eval_{y=y_0} \\
&\phantom{{}={}}+  \sum_i  e^{-4 A} \left[ \left( \frac{\partial V_0}{\partial \phi_i}-2 \phi'_i \right) \dv{ \phi_i}{ y_1} +\frac{1}{2 A'} \left( \phi'_i \dv{\phi'_i}{y_1} -\phi''_i \dv{\phi_i}{y_1} \right) \right] \eval_{y=y_1} \\
&\phantom{{}={}}-  4 e^{-4 A_1} \dv{A_1}{y_1} \tilde{V}_\text{IR} -  4 e^{-4 A_0} \dv{A_0}{y_1} \tilde{V}_\text{UV}.
\end{aligned}
\eeq
The first terms in parentheses in each bracket would vanish if the boundary conditions were imposed.  The other terms in parentheses are Lagrange brackets with respect to the variables $A$ and $y_1$~\cite{LanLifMech}.  Let us see how that is the case.  First, we identify the conjugate momenta to $\phi_i$:
\beq
\pi_i = -\frac{\partial \mathcal{L}}{\partial \phi'_i} = -\sqrt{g} g^{55} \phi'_i = e^{-4 A} \phi'_i,
\eeq
where the minus sign is irrelevant, and chosen for convenience. The Lagrange bracket for the pair $(A,y_1)$ (at some arbitrary $y$-value) is:
\beq
\begin{aligned}
\left\{ A, y_1 \right\}^{\phi_i, \pi_i} &= \sum_i \frac{\partial \pi_i}{\partial A} \frac{\partial \phi_i}{\partial y_1}-\frac{\partial \pi_i}{\partial y_1} \frac{\partial \phi_i}{\partial A}  \\
&=\sum_i \left[ -4 \pi_i  \frac{\partial \phi_i}{\partial y_1} +4 \pi_i  \frac{\partial A}{\partial y_1} \frac{\partial \phi_i}{\partial A} +e^{-4A}  \left( \frac{\partial \phi_i'}{\partial A}\frac{\partial \phi_i}{\partial y_1} -\frac{\partial \phi_i'}{\partial y_1}\frac{\partial \phi_i}{\partial A}  \right) \right].
\end{aligned}
\eeq
The first two terms cancel, and the rest (using $\dd{A} = A' \dd{y}$) gives the remaining terms we have in the brackets in the derivative of the effective potential.  

The final derivative of the potential then is given by:
\beq
\begin{aligned}
\dv{V_\textnormal{eff}}{y_1}  &= e^{-4 A_0} \left( \frac{\partial V_0}{\partial \phi_0}-2 \phi'_0 \right) \dv{\phi_0}{y_1}  +
 e^{-4 A_1} \left( \frac{\partial V_1}{\partial \phi_1}+ 2 \phi'_1 \right) \dv{\phi_1}{y_1}    \\
 &\phantom{{}={}}+ \left\{ A, y_1 \right\}^{\phi_0, \pi_{\phi,0}}-  \left\{ A, y_1 \right\}^{\phi_1, \pi_{\phi,1}}  \\
&\phantom{{}={}}-  4 e^{-4 A_1} \dv{A_1}{y_1} \tilde{V}_\text{IR} -  4 e^{-4 A_0} \dv{A_0}{y_1} \tilde{V}_\text{UV}.
\end{aligned}
\eeq
Translations in $y$ are an example of canonical transformations on the coordinates~\cite{LanLifMech}, and the Lagrange bracket is thus an invariant under shifts in $y$.  Therefore the two Lagrange bracket terms cancel against each other.  The very last term is vanishing by choice of an overall metric scale factor: $\dd A_0/\dd y_1 = 0$.  Imposing the scalar equations of motion on the IR brane causes the second term to vanish.  We are left with Equation~\eqref{eq:dVbydy1}.


\section{A Superpotential Method for Many Scalar Fields}
\label{sec:superpotential}
It is well known that a method exists for constructing static geometries that satisfy all boundary conditions and bulk equations of motion when there is a single scalar field.  The method posits a ``superpotential" function of the scalar field $\phi$, $W(\phi)$.  For a given superpotential, $W$, there is a unique bulk potential given by
\beq
V(\phi) = \frac{1}{8} \left( \frac{\partial W}{\partial \phi} \right)^2 - \frac{\kappa^2}{6} W^2.
\eeq
Taking the $y$-derivative of the above equation and dividing by $\phi'$ gives
\beq
\frac{ \partial V}{\partial \phi} = \frac{1}{4 \phi'} \frac{\partial W}{\partial \phi} \dv{y} \frac{\partial W}{\partial \phi}- \frac{\kappa^2}{3} W \frac{\partial W}{\partial \phi}.
\eeq
From this, we see that if $\phi' = \frac{1}{2} \frac{\partial W}{\partial \phi}$, and $A' = \frac{\kappa^2}{6} W$, this is equivalent to the scalar equation of motion.  The second of these two relations is required by Einstein's equations, as its $y$ derivative must reproduce the relation $A'' = \frac{\kappa^2}{3} \phi'^2$.

If one starts with a given $W$, there must be a relation between $W$ and the boundary potentials for this bulk solution (which is first order) to solve the scalar boundary value problem (which is second order).  Additionally, the boundary potentials are constrained to have the form
\beq
V_{0,1}(\phi) = \pm W( \phi_{0,1} ) \pm W' (\phi_{0,1}) (\phi - \phi_{0,1}) + \text{ higher order polynomial in } (\phi-\phi_{0,1}).
\eeq

With this procedure, one can start with any $W$, and create a static solution to the geometry with any given $y_1$.  This procedure incorporates arbitrary backreaction.  Stability is not guaranteed (the radion may be tachyonic about these points), but this can be checked by solving the mode expansion about the classical background determined by this method.

Using this method, one can also ``go backwards", starting with the (more fundamental) $V(\phi)$, and then integrate the superpotential equation.  It is non-linear, and typically no analytic solutions are possible (with a limited number of exceptions).  The rest of the story proceeds in the same way, but the additional integration constant gives the freedom to match arbitrary boundary potentials (although one must solve for $y_1$, rather than positing it from the get-go).

This procedure can be generalized to multiple scalar fields in the following way.  Take a superpotential $W( \{\phi_i\})$.  This is achieved by defining the bulk potential as
\beq
V = \frac{1}{8} \sum_i \left( \frac{\partial W}{\partial \phi_i} \right)^2 - \frac{\kappa^2}{6} W^2
\eeq
and taking 
\beq
\phi_i' = \frac{1}{2} \frac{\partial W}{\partial \phi_i} \qquad\text{and} \qquad A' = \frac{\kappa^2}{6}.
\eeq
One can now again take the $y$-derivative of the superpotential relation.  The result is a sum over all of the scalar field equations of motion after substituting the above relations, showing that this procedure will generate solutions to the bulk equations of motion, for any given $W$.  The boundary conditions also must be satisfied, which will happen for the following form of the boundary potentials:
\beq
\begin{aligned}
V_{0,1}(\phi) &= \pm W\left( \left\{ \phi_i^{0,1}\right\} \right) \pm \sum_i \frac{\partial W}{\partial \phi_i}\left( \left\{ \phi_i^{0,1}\right\} \right) \left(\phi_i - \phi_i^{0,1}\right) \\
 &\phantom{{}={}}+ \text{ higher order polynomial in } \left(\phi_i-\phi_i^{0,1}\right).
\end{aligned}
\eeq
It is not clear that going in ``reverse" is possible in the same way as in the case of a single scalar field.  That is, it seems unlikely that all solutions that stabilize the geometry with more than one scalar come from superpotentials.

\end{document}